\documentclass[pre,a4paper,superscriptaddress,nofootinbib,twocolumn]{revtex4}

\usepackage{epsfig}
\usepackage{graphicx}
 \usepackage{longtable}
\usepackage{latexsym}

\long\def\symbolfootnote[#1]#2{\begingroup%
\def\thefootnote{\fnsymbol{footnote}}\footnote[#1]{#2}\endgroup}

\newcommand{\ux}{u_{x}}
\newcommand{\uz}{u_{z}}
\newcommand{\Uc}{U_{c}}

\newcommand{\dx}{\partial_{x}}
\newcommand{\dz}{\partial_{z}}
\newcommand{\ddz}{\frac{\partial }{\partial z}}
\newcommand{\ddx}{\frac{\partial }{\partial x}}
\newcommand{\ddxtilde}{\frac{\partial }{\partial \tilde{x}}}

\newcommand{\duxdx}{\frac{\partial u_{x}}{\partial x}}
\newcommand{\duxdxx}{\frac{\partial^{2} u_{x}}{\partial x^{2}}}
\newcommand{\duxdz}{\frac{\partial u_{x}}{\partial z}}
\newcommand{\duxdzz}{\frac{\partial^{2} u_{x}}{\partial z^{2}}}
\newcommand{\duzdx}{\frac{\partial u_{z}}{\partial x}}
\newcommand{\duzdz}{\frac{\partial u_{z}}{\partial z}}

\newcommand{\dTdx}{\frac{\partial T}{\partial x}}
\newcommand{\dTdz}{\frac{\partial T}{\partial z}}

\newcommand{\dsdx}{\frac{\partial s}{\partial x}}
\newcommand{\dsdz}{\frac{\partial s}{\partial z}}

\newcommand{\drhodx}{\frac{\partial \rho}{\partial x}}
\newcommand{\drhodz}{\frac{\partial \rho}{\partial z}}

\newcommand{\detadz}{\frac{\partial \eta}{\partial z}}

\newcommand{\dPdT}{\left( \frac{\partial P}{\partial T} \right)_{\rho}}
\newcommand{\dsdT}{\left( \frac{\partial s}{\partial T} \right)_{\rho}}
\newcommand{\dsdrho}{\left( \frac{\partial s}{\partial \rho} \right)_{T}}

\newcommand{\lc}{\ell_{c}}

\newcommand{\xtilde}{\tilde{x}}
\newcommand{\ztilde}{\tilde{z}}
\newcommand{\ztildex}{\tilde{Z}}

\newcommand{\uxtilde}{\tilde{u}_{x}}
\newcommand{\uztilde}{\tilde{u}_{z}}
\newcommand{\Psitilde}{\tilde{\Psi}}
\newcommand{\Thetatilde}{\tilde{\Theta}}

\newcommand{\rhotilde}{\tilde{\rho}}
\newcommand{\etatilde}{\tilde{\eta}}
\newcommand{\betatilde}{\tilde{\beta}}
\newcommand{\Lambdatilde}{\tilde{\Lambda}}
\newcommand{\cptilde}{\tilde{c}_{P}}

\newcommand{\duxdxtilde}{\frac{\partial \tilde{u}_{x}}{\partial \tilde{x}}}
\newcommand{\duxdxxtilde}{\frac{\partial^{2} \tilde{u}_{x}}{\partial \tilde{x}^{2}}}
\newcommand{\duxdztilde}{\frac{\partial \tilde{u}_{x}}{\partial \tilde{z}}}
\newcommand{\duxdzztilde}{\frac{\partial^{2} \tilde{u}_{x}}{\partial \tilde{z}^{2}}}
\newcommand{\duzdxtilde}{\frac{\partial \tilde{u}_{z}}{\partial \tilde{x}}}
\newcommand{\duzdztilde}{\frac{\partial \tilde{u}_{z}}{\partial \tilde{z}}}
\newcommand{\detadztilde}{\frac{\partial \tilde{\eta}}{\partial \tilde{z}}}

\newcommand{\detadtheta}{\left( \frac{\partial \tilde{\eta}}{\partial \tilde{\Theta}}\right)_{\tilde{\rho}}}
\newcommand{\detadrho}{\left( \frac{\partial \tilde{\eta}}{\partial \tilde{\rho}} \right)_{\tilde{\Theta}}}

\begin{document}

\title{Non-Oberbeck-Boussinesq effects in turbulent thermal convection in 
ethane close to the critical point}

\author{Guenter Ahlers}
\affiliation{Department of Physics and iQCD, University of California, Santa Barbara, CA 93106}

\author{Enrico Calzavarini}
\affiliation{Department of Applied Physics and J. M. Burgers Centre for 
Fluid Dynamics, University of Twente, 7500 AE Enschede, The Netherlands}

\author{Francisco \surname{Fontenele Araujo}}
\affiliation{Department of Applied Physics and J. M. Burgers Centre for 
Fluid Dynamics, University of Twente, 7500 AE Enschede, The Netherlands}

\author{Denis Funfschilling}
\affiliation{LSGC CNRS - GROUPE ENSIC, BP 451, 54001 Nancy Cedex, France}

\author{Siegfried Grossmann}
\affiliation{Department of Physics, Philipps-University of Marburg, Renthof 6, 
D-35032 Marburg, Germany}

\author{Detlef Lohse}
\affiliation{Department of Applied Physics and J. M. Burgers Centre for 
Fluid Dynamics, University of Twente, 7500 AE Enschede, The Netherlands}

\author{Kazuyasu Sugiyama}
\affiliation{Department of Applied Physics and J. M. Burgers Centre
for Fluid Dynamics, University of Twente, 7500 AE Enschede, The Netherlands}

\begin{abstract}
As shown in earlier work (Ahlers {\it et al.}, 
J. Fluid Mech. {\bf 569}, 409 (2006)), non-Oberbeck Boussinesq (NOB) corrections
to the center temperature in turbulent Rayleigh-B\'enard convection
{\it in water} and also {\it in glycerol}
are governed by the temperature dependences of 
the kinematic viscosity and the thermal diffusion coefficient. 
If the working fluid is {\it ethane close to the critical point} 
the origin of non-Oberbeck-Boussinesq corrections is very different, 
as will be shown in the present paper.
Namely, the main origin of NOB corrections then lies in
the strong temperature dependence of the isobaric thermal expansion coefficient $\beta(T)$. More 
precisely, it is the nonlinear $T$-dependence of the density $\rho(T)$ in the buoyancy force 
which causes another type of NOB effect.
We demonstrate that through a combination of experimental, numerical, and theoretical
work, the latter in the framework of the extended Prandtl-Blasius
boundary layer theory developed by Ahlers {\it et al.}, 
J. Fluid Mech. {\bf 569}, 409 (2006). The latter comes to its limits, if the 
temperature dependence of the thermal expension coefficient $\beta(T)$ is significant. The new measurements reported here cover the ranges $2.1 \alt Pr \alt 3.9$ and $5\times 10^9 \alt Ra \alt 2\times 10^{12}$ and are for cylindrical samples of aspect ratios 1.0 and 0.5.
\end{abstract}

\date{\today}


\maketitle

\section{Introduction}
Fluid motion in the presence of temperature gradients is an important
phenomenon in nature and industrial processes. Among the many examples  
are oceanic streams, cloud motions, and gusts of wind that can be felt 
on a human scale.
The paradigmatical system for such thermally driven flows is 
the Rayleigh-B\'enard (RB) setup: a fluid-filled container heated 
from below and cooled from above. In this classical problem, the flow  
is determined by the scale and 
geometry of the container, the material properties of the 
working fluid, and the top-down temperature difference $\Delta \equiv 
T_{b} - T_{t} > 0$. 
In the last two decades, 
considerable progress has been achieved
in our understanding of global and local properties and the flow organization
of turbulent RB convection, 
through a combination of experimental (see e.g.\ 
\cite{cas89,sig94,cio97,cha97,xu00,nie00,cha01,ahl01,qiu01b,kad01,xia02,roc02,nie03,fun04,bro05b,nik05,nie06,xia03,sha03,roc04,sun05,bro06,nie06b,pui07}),
numerical (see e.g.\ \cite{ker96,ben98,ker00,ver99,ver03,loh03,ver04,ama05,shi06,str06,kun08}), and theoretical work
(see e.g.\ \cite{gro-all,ben05,bro07}). 

The temperature difference in a RB cell can be increased in a controlled way.
 However, in principle
the transport coefficients of the fluid can depend on the local temperature and density and
thus vary across the height $L$ of the container. 
Since space-dependent properties of such kind are undesirable in 
first instance, one tends to restrict the convection regime to 
sufficiently small intervals of $\Delta$. But even so, further 
simplifications are progressively required in the analysis of RB 
convection. In this spirit, a standard approximation due to Oberbeck 
\cite{obe79} and Boussinesq \cite{bou03} assumes that 
(see also \cite{ll87,cha81}): 

\begin{itemize}

\item[OB.1] 
The dynamic viscosity $\eta$, the thermal conductivity $\Lambda$, 
the thermal expansivity $\beta$, and the isobaric specific heat $c_{P}$ 
are constant throughout the fluid.

\item[OB.2] 
Density variations are taken into account only in the buoyancy force term. 

\item[OB.3] 
The temperature dependence of the density $\rho$   
is linearized in the buoyancy force as: 
\begin{equation}
\label{density/linear}
\rho(T)\;\; = \;\; \rho_m \;-\; \rho_{m}\,\beta_m\,(T - T_m),
\end{equation}
where $T_{m} \equiv (T_{b}+T_{t})/2$ is the arithmetic mean temperature between 
the plates and $X_{m} = X(T_{m})$ denotes the fluid property $X$ 
evaluated at $T_{m}$.
\end{itemize}

Next to the aspect ratio, within
 the OB approximation two dimensionless parameters 
characterize the RB flow: The Prandtl number 
$Pr~\equiv~\nu_{m}/\kappa_{m}$ follows from the ratio between the 
kinematic viscosity $\nu_m\equiv\eta_m/\rho_m$ 
and the thermal diffusivity $\kappa_m\equiv\Lambda_m/(\rho_mc_{P,m})$. 
The dimensionless thermal driving can be conveniently represented 
by the Rayleigh number $Ra~\equiv~\beta_m g L^{3} \Delta/(\nu_{m} \kappa_{m})$, 
where $g$ denotes the gravitational acceleration.

The manner in which high Rayleigh numbers are achieved is 
crucial for the emergence of non-Oberbeck-Boussinesq effects (NOB). 
Since turbulent convection may involve spatiotemporal changes   
in the fluid properties, considerable efforts have been devoted to 
the identification of dominating sources of NOB effects. In liquids like 
water \cite{ahl06} and glycerol \cite{zha97}, \cite{sug07}, for example, NOB 
effects are dominated by deviations from (OB.1) since the viscosity 
strongly decreases with temperature. On the other hand, when the working 
fluid is gaseous ethane \cite{ahl07}, deviations from (OB.1) and (OB.2) 
lead to NOB effects stronger than those in the aforementioned liquids.

In the present study, we shall focus on deviations from (OB.3) by considering 
the nonlinear temperature dependence of the buoyancy force. In particular, 
ethane close to its critical point \cite{fri91b} is chosen as the working 
fluid and the temperature $T_c$ in the center of the container is measured 
as indicator of NOB effects. 

There are two possibilities to characterize the physics beyond condition OB.3 as described by 
eq. (\ref{density/linear}), if the density $\rho(T)$ has a strong nonlinear $T$-dependence. 
First, one can introduce a $T$-dependent thermal expansion function $\hat{\beta}(T)$ instead 
of $\beta_m$, defined in terms of the density $\rho(T)$ by
\begin{equation} \label{hatbeta} 
\rho(T) \equiv \rho_m - \rho_m (T - T_m) \hat{\beta}(T) . 
\end{equation}
Second, one refers to the common isobaric thermal expansion coefficient
$\beta(T)$, defined as usual by
\begin{equation} \label{beta}
\beta(T) \equiv - \frac{1}{\rho(T)} \frac{\partial \rho (T)}{\partial T} |_P ,
\end{equation}
which now is temperature dependent. Both are related by $\beta(T) = - \partial ~\mbox{log} [1 - (T-T_m) 
\hat{\beta}(T)] ~/ ~\partial T$. Under the condition OB.3 of linear $T$-dependence of $\rho$ the 
thermal expansion function is constant, $\hat{\beta} = \beta_m$, while the expansion coefficient $\beta(T)$ 
is given by $\beta(T) = \beta_m / [1 - (T-T_m)\beta_m]$, still depending on temperature. Of course all 
three coincide at $T_m$. An advantage of considering the thermal expansion coefficient $\beta(T)$ is that 
it is a well defined thermodynamic derivative. The advantage of $\hat{\beta}(T)$ on the other hand is that it 
immediately reflects the nonlinear $T$-dependence of $\rho(T)$. Also $\beta(T)$ refers to a single
thermodynamic state and describes the local $T$-slope on an isobar (normalized by the local density), 
while $\hat{\beta}(T)$ refers to a pair of states, namely to the reference state $T_m$ in addition 
to $T$ and describes the secant to the $\rho(T)$-curve (normalized by the reference density). 
$\hat{\beta}(T)$ will therefore in general vary less with $T$ than $\beta(T)$.  

It will turn out that it is the 
significantly different $T$-dependence of $\hat{\beta}(T)$ (or $\beta(T)$) on the two sides of the critical 
isochore of ethane, which leads to opposite shifts of the center (bulk) temperature 
$T_c$, yielding $T_c < T_m$ on the gas-like (i.e. high-temperature, see Fig.~\ref{phase_diagram} below) side and $T_c > T_m$ on the liquid-like (low-temperature) side.
On the gas-like side $\hat{\beta}(T)$ increases from bottom to top and on the liquid-like side it 
decreases.     
  
Our approach consists of three stages: 
boundary layer (BL) theory, experiments, and direct numerical simulations 
(DNS). First, we address in section \ref{section/blt} an extension of
boundary layer theory that considers deviations from (OB.1) and (OB.2). 
Even though the buoyancy force is \textit{not} included in the BL equations 
(only the \textit{longitudinal} momentum is taken into account here), we compute 
$T_{c}(\Delta)$ for several pressures $P_{m}$. Then, experimental measurements of 
$T_{c}(\Delta)$ are presented in section \ref{section/exp_results} and compared 
with BL results in section \ref{section/comparison}. Given the significant 
discrepancies between part of them, we address in section \ref{section/dns} 
direct numerical simulations that explicitly consider deviations from (OB.3). In particular, 
for $T_m = 27^{\circ}$~C and $P_{m} = 51.72$ bar, it is shown that NOB 
effects in ethane are dominated by the nonlinear dependence of the buoyancy 
force on temperature. Finally, our conclusions are summarized in section 
\ref{section/conclusions}. Appendices \ref{section/appendix} and \ref{appb} are devoted 
to the derivation of the boundary layer equations with variable transport coefficients and
Appendix \ref{appc} compiles the Nusselt number corrections for the real and various hypothetical 
ethane-like fluids in a table.

\section{Boundary-layer theory}
\label{section/blt}

A central aspect in Rayleigh-B\'enard convection is the understanding 
of the boundary layers formed along the bottom and top plates. 
Though they preserve a laminar character for $Ra$ $\leq 10^{12}$, 
their instabilities impact the Nusselt number $Nu$ (the effective heat 
flux relative to thermal conduction $\Lambda_{m}\Delta/L$) \cite{cil96,ahl06b}. 
As reported in references \cite{ahl07,ahl06,zha97}, BL flows of this 
nature are significantly influenced by the coupling between the fluid 
properties and the temperature gradient across the container. 
In particular, it was shown that NOB effects on $T_{c}$ can be reasonably  
described by extending the Prandtl-Blasius boundary layer theory 
\cite{sch00,ste64}. Next we review such 
an extension and further assess its intrinsic limitations.

Assume that the density $\rho$, the temperature $T$, and the velocity $\mathbf{u}$ 
are stationary fields, which depend only on the longitudinal $x$ and transverse $z$ 
coordinates. Then, under the boundary-layer approximation, we write the continuity 
and the $x$-momentum equations as (see also 
appendix \ref{appendix/viscous-bl/2d}):
\begin{eqnarray}
\label{continuity}
\ddx\{\rho \ux \} + \ddz\{\rho \uz \} 
& \;\;=\;\; & 0,\\
\label{Prandtl}
\rho\,\left\{\ux \; \duxdx + \uz \; \duxdz \right\}
& \;\;=\;\; &
\ddz \left\{ \eta \, \duxdz \right\}.
\end{eqnarray}
Here $z$ measures the vertical distance from the bottom or top plates, respectively, 
the velocity components at $z=0$ are subject to no-slip boundary 
conditions: $\ux(x,0) = 0$ and $\uz(x,0) = 0$. Moreover, in the bulk 
of the flow, $\ux$ is supposed to match the large scale 
wind velocity $U_{c}$ in the center (bulk) of the RB sample 
\cite{kri81}, 
i.e., $\ux(x,\infty) = U_c$.
Note that within the BL theory we cannot calculate $U_c$; here we only have
to assume that it is the same close to the top and the bottom BL which is
supported by our numerical simulations reported in section V.

In the same spirit, the temperature field $T(x,z)$ is 
governed by (cf. appendix 
\ref{appendix/energy/gas}):
\begin{eqnarray}
\frac{\gamma}{\rho\,c_P}
\ddz\left\{\Lambda\,\dTdz \right\}
&\;=\;&
\nonumber
\ux\,\dTdx \;+\; \uz \, \dTdz \\
\label{temperature/NOB/gas}
&&
\;+\;
\frac{\gamma - 1}{\beta}
\left\{ \duxdx + \duzdz \right\}
,
\end{eqnarray}
where $\gamma \equiv c_P/c_V$ is the ratio between the isobaric and isochoric 
specific heats and $\beta = - \rho^{-1} \partial \rho / \partial T$ denotes the isobaric thermal 
expansion coefficient. At the plates $T(x,0) = T_{b,t}$ and in the bulk (center) of the flow 
$T(x,\infty) = T_{c}$. 

The coupling between the bottom and top boundary layers is determined (cf. \cite{ahl06}) by 
the heat fluxes $Q_{b,t}$ through the plates, considered to be equal:
\begin{equation}
\label{heat-fluxes}
Q_{b}
\;\;=\;\;
- \Lambda_{b}\,\left. \dTdz \right|_{b}
\;\;=\;\;
- \Lambda_{t}\,\left.\dTdz \right|_{t}
\;\;=\;\;
Q_{t}.
\end{equation}
This condition establishes an implicit dependence of the 
center temperature $T_c$ on the heat fluxes $Q_b = Q_t = Q$. 
Note again that both the dynamic viscosity $\eta$ and the heat
conductivity $\Lambda$ depend on both temperature and density, i.e.,
$\eta(T,\rho)$ and
$\Lambda(T,\rho)$.
Before addressing  
the technicalities around the (numerical) integration of equations 
(\ref{continuity})--(\ref{heat-fluxes}), we shall benefit from a key 
argument in boundary-layer theory: Prandtl's self-similar ansatz.

\subsection{Self-similarity}

Because of the y-independence, assumed in the Prandtl BL theory, the boundary layer flow is 
mathematically a 2D flow. Therefore the   
system of partial differential equations (\ref{continuity})--(\ref{temperature/NOB/gas}) for the BL flow 
can be reduced to ordinary differential equations (ODEs) by introducing a stream function $\Psi$. We do this 
differently
 from the usual procedure by including in its definition the density in order to automatically fulfil
the continuity equation by construction.
\begin{eqnarray}
\label{rhotilde.ux}
\rhotilde\,\ux 
&\;=\;&
\;\frac{\partial \Psi}{\partial z},\\
\label{rhotilde.uz}
\rhotilde\,\uz 
&\;=\;&
-\; \frac{\partial \Psi}{\partial x},\,
\end{eqnarray}
where $\tilde \rho \equiv \rho/\rho_m$ is the density nondimensionalized 
with $\rho_m = \rho(T_m, P_m)$. Apparently the continuity equation 
automatically follows from (\ref{rhotilde.ux})--(\ref{rhotilde.uz}).  
Next, cf. Appendix \ref{appb}, we may introduce a self-similarity variable 
$ 
\ztildex  
\equiv
z/ \lc(x)
$ 
and a similarity function 
$\Psitilde(\ztildex) = {\Psi(x,z)}/{(\lc (x)\; \Uc)}$,
such that $\lc(x)= \sqrt{x \nu_{m}/ \Uc}$. 
Thus the velocity components are
\begin{equation}
\label{ux/Psi/NOB/gas}
\ux 
\;=\; 
\Uc\,\frac{\Psitilde\,'}{\rhotilde},
\qquad
\uz 
\;=\; 
\frac{\nu_{m}}{2\lc}
\left\{ 
\ztildex\,\frac{\Psitilde\,'}{\rhotilde} - \frac{\Psitilde}{\rhotilde}
\right\},
\end{equation}
with boundary conditions    
$\Psitilde(0) = 0 = \Psitilde\,'(0)$ 
and 
$\Psitilde\,'(\infty)=\rhotilde_c$.

In terms of (\ref{ux/Psi/NOB/gas}), the viscous BL equation 
(\ref{Prandtl}) can be written as:
\begin{eqnarray}
\label{Blasius}
0 &= & \etatilde 
\Psitilde\,''' 
+ 
\left\{
\frac{1}{2}\Psitilde
+ 
\etatilde\,'
- 
2 \frac{\rhotilde\,'}{\rhotilde}\,\etatilde
\right\}\Psitilde\,''
\nonumber \\ 
&+&
\left\{
-\frac{1}{2}\frac{\rhotilde\,'}{\rhotilde} \Psitilde
+
\left[ 2\left(\frac{\rhotilde\,'}{\rhotilde}\right)^2 
       - \frac{\rhotilde\,''}{\rhotilde}\right]\etatilde
-
\frac{\rhotilde\,'}{\rhotilde} \etatilde\,'
\right\}
\Psitilde\,'
.
\end{eqnarray}
Here  
$\tilde \eta \equiv \eta/\eta_m$ 
is the dimensionless viscosity, 
whose $\ztildex$-dependence $\etatilde\,'$ is given by  
\[
\etatilde\,' 
\;=\;
\detadtheta\,\Thetatilde\,'
+
\detadrho\,\rhotilde\,',
\] 
where 
$\Thetatilde 
~\equiv~ 
(T - T_{t})/\Delta
$ 
denotes the dimensionless temperature.

Next, assuming that the pressure $P_m$ is constant 
throughout the fluid, one finds
\begin{equation}
\label{rhotilde/slope}
\rhotilde\,' 
\;\;=\;\; 
-\rhotilde\, \betatilde\, \Thetatilde\,',
\end{equation}
with $\tilde \beta \equiv \beta\,\Delta$. The boundary conditions 
at the respective walls are 
$\rhotilde(0)~=~\rhotilde_{b,t}$, 
$\rhotilde\,'(0)~=~-\rhotilde_{b,t}\,\betatilde_{b,t}\,\Thetatilde\,'_{b,t}$,
and 
$\rhotilde(\infty)~=~\rhotilde_{c}$. 

Finally, we also write the temperature equation (\ref{temperature/NOB/gas}) 
in self-similar form as (see Appendix \ref{appendix/thermal-bl/gas})
\begin{equation}
\label{theta-eq}
\Lambdatilde\,\Thetatilde\,'' 
\;+\;
\left\{
\frac{1}{2}\,\cptilde \, \Pr\, \Psitilde
\;+\;
\Lambdatilde\,'
\right\}\,\Thetatilde\,'
\;\;=\;\;
0  ,
\end{equation}
where $\tilde \Lambda \equiv \Lambda/\Lambda_m$ and $\tilde c_P \equiv c_P/c_{P,m}$.  
Equation~(\ref{theta-eq}) is subject to $\Thetatilde(0) = \Thetatilde_{b,t}$ 
and $\Thetatilde(\infty) = \Thetatilde_{c}$.

\subsection{Results}

\begin{figure}[!ht]

  \begin{center}

    \includegraphics*[height=7cm]{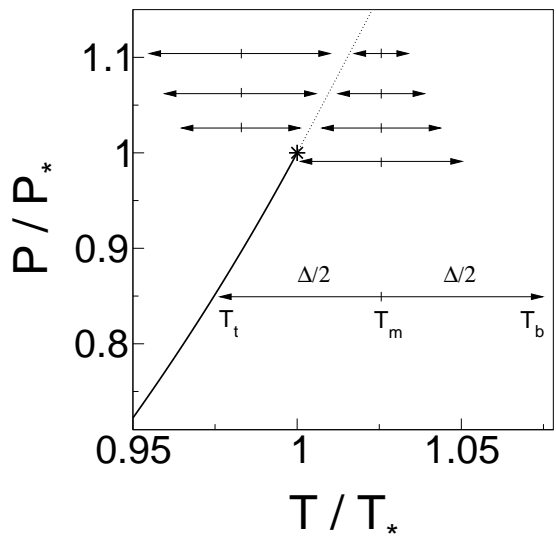}
    \caption{
             \label{phase_diagram}
             Pressure-temperature plane in reduced units. 
             Star: critical point of ethane ($T_* = 32.18^\circ$C, $P_* = 48.718$ bar).  
             Heavy line: liquid-vapor coexistence curve. 
             Dotted line: critical isochore. 
             The horizontal arrows show the maximum temperature intervals $\Delta$, 
             centered at $T_m = 27\,^\circ$C (left) and $T_m = 40\,^\circ$C (right). 
             The pressures are $P_m/P_*= 0.849$, 0.991, 1.026, 1.062, and 1.104 
             (bottom to top).
            }

  \end{center}

\end{figure}

The coupled ODEs (\ref{Blasius})--(\ref{theta-eq}) with the respective boundary 
conditions and the heat-flux conservation (\ref{heat-fluxes}) are solved numerically 
with a shooting method \cite{pre86}. The integration domain is restricted to 
$\Delta$-intervals where the transport properties are concave/convex functions 
of the temperature. In particular, we have chosen ethane as the working fluid 
since its properties are known very well \cite{fri91b}, even close to its critical 
point $(T_{*}, P_{*},\rho_{*})$ [see figure \ref{phase_diagram}]. All material properties 
$\eta, \Lambda, \rho, \beta$, and $c_p$ are implemented in their full dependence on $T$
In this manner, the computation of temperature and density profiles does not involve any fit 
parameter.

\subsubsection{Vertical profiles}
An insight into the structure of the BLs can be achieved
 by studying typical profiles along 
the $z$-direction. To describe them, let us consider a representative case 
in which the pressure is fixed at $P_m = 0.849 \cdot P_{*}$, the mean temperature 
at $T_m = 40\,^{\circ}$C, and the thermal difference between the plates at $\Delta = 15$ K.  

\begin{figure}[!htb]
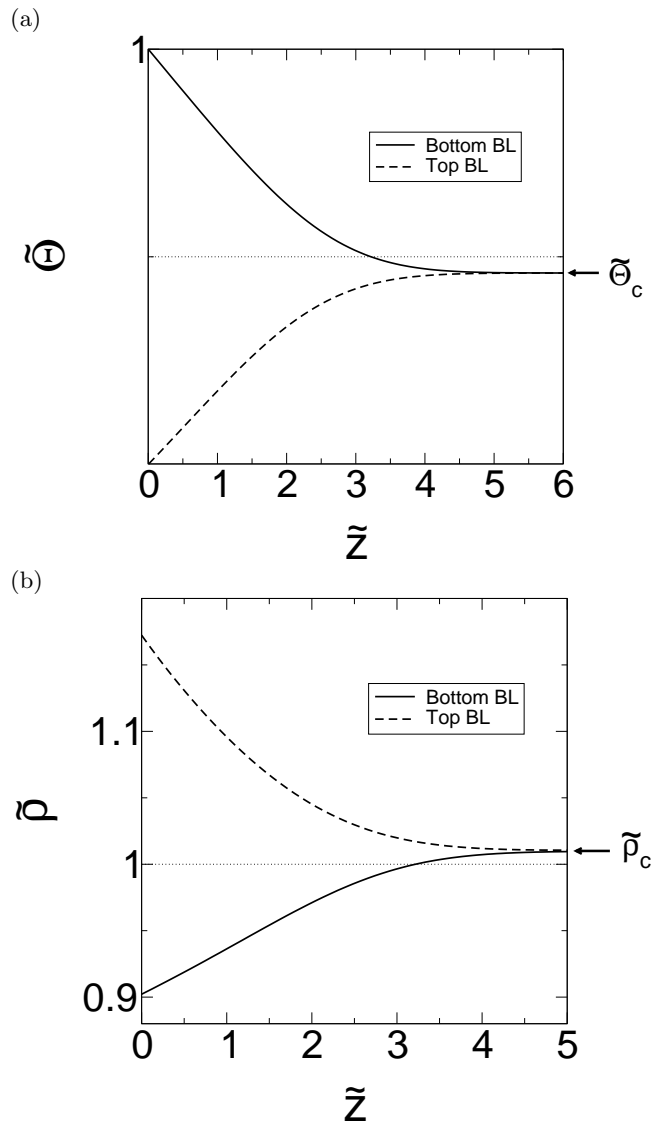

\begin{center}
$\begin{array}{c}
\multicolumn{1}{l}{\mbox{(a)}}\\
\includegraphics*[height=7cm]{Figures_Fontenele_Theta_profile_15_fig2.eps}\\
\multicolumn{1}{l}{\mbox{(b)}}\\
\includegraphics*[height=7cm]{Figures_Fontenele_rho_profile_15_fig3.eps}
\end{array}
$
\end{center}
\caption{
\label{fig/profiles}
(a) Temperature and (b) density profiles at $P_m = 0.849 \cdot P_{*}$, 
$T_m = 40^{\circ}$ C and $\Delta = 15$ K. The thermal slope thicknesses 
at the bottom or top $\lambda^{sl}_{b,t} / L = a_{b,t} / \sqrt{\nu_m / L U_c}$ are seen to have
prefactors of about $a_b \approx 2.8$ and $a_t \approx 2.5$, i.e., $\lambda^{sl}_b > \lambda^{sl}_t$.  
}
\end{figure}

In figure \ref{fig/profiles}, the temperature $\Thetatilde$ 
and density $\rhotilde$ are plotted as functions of the similarity variable 
$\ztildex$. As shown in panel \ref{fig/profiles}a, 
the center temperature $\Thetatilde_{c}$ is \textit{smaller} than the mean 
temperature $\Thetatilde_m = 0.5$, clearly indicating a top-down symmetry 
breaking. Such symmetry breaking is also reflected 
in the density profiles shown in panel \ref{fig/profiles}b, since the center 
density $\rhotilde_{c}$ is \textit{larger} than the mean density $\rhotilde_{m} = 1$. 
Notwithstanding the pronounced curvatures in $\Thetatilde(\ztildex)$ and 
$\rhotilde(\ztildex)$, we shall restrict our attention to the 
\textit{asymptotic} 
value $\Thetatilde_{c}$ as a convenient indicator of NOB effects.

\subsubsection{Center temperature} 

To compute the difference $T_c-T_m$ as a function of $\Delta$, we have chosen    
a particular set of isobars in the phase diagram of ethane. As shown in 
figure \ref{phase_diagram}, our selection of $\Delta$-intervals falls into  
two classes: (i) those intervals centered at $T_m = 27\,^{o}$C and 
(ii) those centered at $T_m = 40\,^{o}$C.

\begin{figure}[!h]
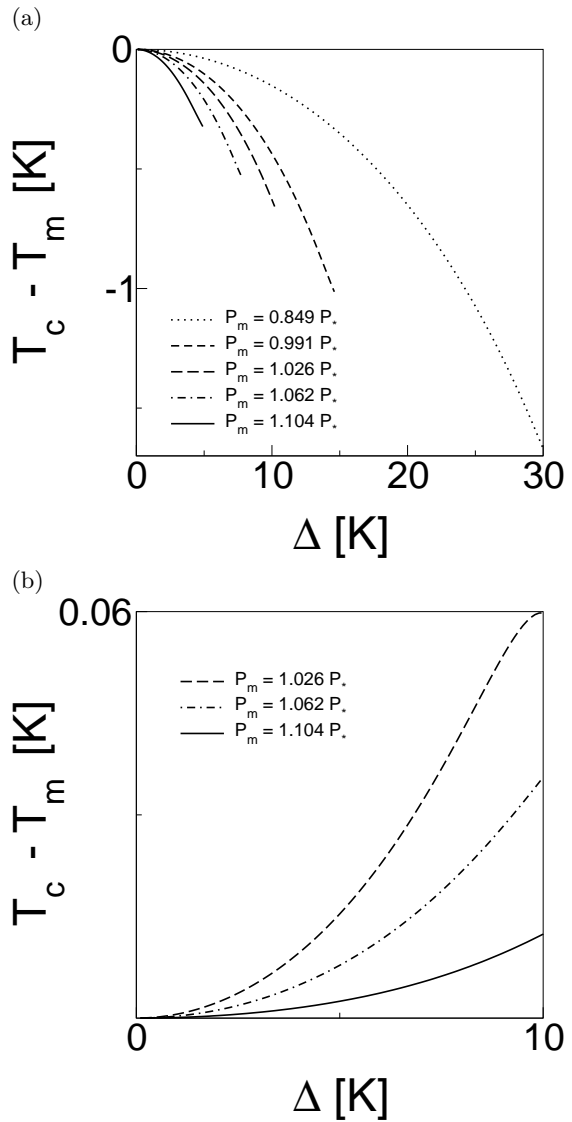


  \begin{center}

   $\begin{array}{c}    

    \multicolumn{1}{l}{\mbox{(a)}}\\
    \includegraphics*[height=7cm]{Figures_Fontenele_T_c-T_m---Theory_at_40_fig5.eps}\\
    \multicolumn{1}{l}{\mbox{(b)}}\\
    \includegraphics*[height=7cm]{Figures_Fontenele_T_c-T_m---Theory_at_27_fig6.eps}
   
   \end{array}
   $

    \caption{
             \label{figure/Tc-bl} 
             Deviation $T_c-T_m$ as function of $\Delta$, 
             for (a) $T_{m} = 40\,^{\circ}$C and 
             (b) $T_{m} = 27\,^{\circ}$C.
            }

  \end{center}

\end{figure}

As for the latter (the more gaseous case), 
figure \ref{figure/Tc-bl}a shows that the center 
temperature is a decreasing function of $\Delta$. The top-down 
symmetry of the flow is broken in such a way that the top boundary layer 
tends to become thinner than its bottom counterpart, 
eventually leading to a temperature reduction
in the center of the flow.
Though this result has 
been originally reported and explained 
already in reference \cite{ahl07}, we briefly mention 
it here for completeness of discussion.

Focussing now on the class of $\Delta$-intervals centered at $T_{m} = 27^{\circ} \,$C, 
figure \ref{figure/Tc-bl}b shows that the center temperature becomes larger than 
the mean temperature between the plates. Such NOB effect 
is different to what 
we found in reference \cite{ahl07} and repeated in figure \ref{figure/Tc-bl} a, where we have focused on the more
gas-like case. 

To understand this we argue that the intervals under 
consideration (centered at $T_{m} = 27^{\circ} \,$C) now
correspond to a region of the phase diagram, where the material properties of ethane 
behave more similar to those of the liquid phase. NOB effects in classical liquids (such as water and glycerol) 
were already discussed in references \cite{ahl06,zha97,sug07}. One of our aims in 
the present work is to further assess the differences in the NOB effects between the more liquid-like versus 
the more gas-like fluids, see Sec.\ \ref{section/exo_results}. It will turn out that this will  
show us the limitations of boundary-layer theory, see Sec.\
 \ref{section/comparison}. 
To this end, we first consider now additional experimental details on
the $\Delta$ dependence of the center temperature $T_{c}(\Delta)$.

\section{Experiment}\label{section/exo_results}

\subsection{Apparatus}
\label{section/apparatus}

\begin{figure}
\centerline{\psfig{file=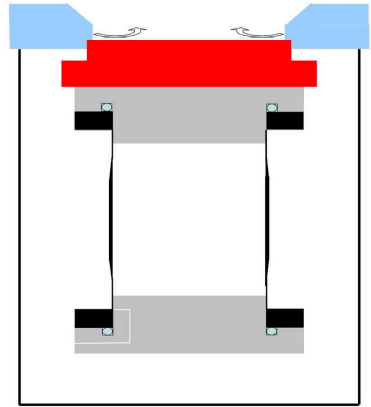,width=2in}}
\caption{Schematic diagram (approximately to scale for $\Gamma = 1$) of the high-pressure sample cell surrounded by the can containing ambient air and foam.}
\label{fig:sample} 
\end{figure}

The apparatus was described in detail before in Ref.~\cite{ACBS94},
where a schematic diagram is shown in Fig.\ 2 . Here we give a brief description and details specific to the present high-pressure sample cell shown in Fig.~\ref{fig:sample}. Working from the inside out, the sample cell was surrounded by a can containing ambient air. The air space inside the can was filled with low-density open-pore foam to prevent convection outside the sample. The maximum possible diameter of the sample top plate was 10 cm, allowing for inside sample diameters typically up to about $D = 8$ cm. The entire apparatus was of sufficient length to accommodate a sample with $L \simeq 16$ cm ($\Gamma \simeq 0.5$). Heat was applied at the sample bottom by a metal-film heater covering the entire active bottom-plate area uniformly.

The top plate was cooled by a circulating water bath. The water was cooled when passing through a heat exchanger external to the main apparatus which in turn was cooled by a separate water circuit driven by a Neslab or Lauda refrigerated circulator with a temperature stability of 0.01$^\circ$C. Just before entering the apparatus, the water was heated by a heater consisting of about 25 m of teflon-insulated AWG30 (0.5 mm dia) copper wire, stuffed into the inlet line and thus immersed in the water. The large contact area between the water and the heater wire provided excellent heat exchange and uniform heating of the water. The heater was computer controlled in a feedback loop with a thermometer located in the top plate of the sample cell. The bath-temperature stability achieved in this way was a few tenths of a milli-Kelvin. The water entered the bottom center of the apparatus, flowed upward through an annular channel around the can,  and was distributed over the top plate by a set of jets. With this arrangement the entire can was kept at the top-plate temperature and parasitic heat loss from the side wall and the bottom plate due to conduction through the air/foam as well as by radiation was reduced to a level that was negligible compared to the heat transport by the convecting fluid. 

After cooling the top plate the water returned through an annular channel located just outside of and mildly insulated from the incoming channel. Since the water, while cooling the top plate,  was never heated by more than a few mK, the returning water provided an excellent adiabatic thermal shield at the top-plate temperature, thus stabilizing the interior temperatures and preventing significant variations in time of the parasitic heat losses from the bottom plate. The entire apparatus sat on a chlorinated poly (vinyl chloride) (CPVC) base plate with appropriate channels and feed-throughs to accommodate the water circuit, the electrical leads, and the fill capillary going to the sample. 

One of two  high-pressure sample-cells (see Fig.~\ref{fig:sample}) was installed inside the can. It had the shape of a cylinder with $D = 7.63$ cm. One cell had an internal length $L = 7.62$ cm, corresponding to an aspect ratio $\Gamma \equiv D/L = 1.00$. Another one had $L = 15.24$ cm, yielding $\Gamma = 0.500$. The top and bottom plates consisted of thick  copper disks. Each of the two plates had an anvil, 1.59 cm thick, of diameter essentially equal to $D$, that was a close slide fit in the type 4340 steel side wall. After machining, the side wall was heat treated at 830$^\circ$C and oil quenched. This procedure is expected to lead to a tensile strength of about 13 kbars. 
The side wall had a flange at each end, of thickness 0.95 cm. Each flange was bolted (not shown in the figure) and ``O"-ring-sealed to one of the copper end plates. A top and bottom thin section of the side wall had a thickness of 0.051 (0.076) cm for the $\Gamma = 1$ ($\Gamma = 0.5$) cell. This thin section overlapped the copper anvils and extended into the sample region by 0.95  cm. Connecting the thin section was a central section of wall thickness 0.15 cm that provided enhanced strength; since the turbulent system contained only a very small thermal gradient in its center, the thicker wall section did not significantly enhance the wall heat-transport. The sample entered the bottom copper plate through a capillary from the side, and then proceeded through a very small hole  (shown in white on the left side of the figure) into the gap between the bottom-plate anvil and the side wall.

The sample was connected to a manifold through a capillary. Also connected to the manifold was a separate pressure-regulation volume of 600 (1000) cm$^3$ in the case of the $\Gamma = 1$ ($\Gamma = 0.5$) cell that could be heated above the ambient temperature  by a heater wrapped around its outside. The temperature of this ``hot volume" was controlled in a feedback loop with a pressure gage \cite{mue76}. The pressure stability typically was better that one milli-bar. The entire system was designed to safely withstand pressures up to 60 bars.
For the pressure measurements we used a Paroscientific model 745 pressure standard with an accuracy or 80 ppm (about 6 milli-bars) and a resolution of 1 ppm (about 70 micro-bars) of full scale.

A substantial fraction of the heat current passed through 
the side wall. This current was measured for the evacuated cell and subtracted from all 
other measurements; but as was recognized some time ago (\cite{ahl00,RCCHS01}), this is 
not an adequate procedure because of the height-dependent temperature gradients that 
prevail in the wall when the cell contains turbulently convecting fluid. We did not 
attempt a correction for this non-linear side-wall effect in the present case because 
we do not believe that a reliable correction is possible when the side-wall conductance 
is large. For this reason our values of ${Nu}$ under OB condition are about 25\% larger 
than other measurements at similar Prandtl numbers (\cite{nik05}). However, we believe 
that the deviations of $Nu$ and of $T_c$  from their Boussinesq values, which (as we 
shall see below in Sect.~\ref{sec:Gamma}) depend primarily on the nature of the top and 
bottom boundary layers rather than on the fluid interior, were obtained reliably.

The top and bottom temperatures $T_t$ and $T_b$  were determined from the average 
of six thermistors imbedded close to the fluid in each of the top and bottom plates (\cite{bro05}).  These thermistors were calibrated against a platinum-resistance thermometer purchased from Hart Scientific. This thermometer was supplied with a calibration, accurate to $\pm7$ mK, on the ITS90 temperature scale.  The average temperature readings were
used to obtain $\Delta = T_b-T_t$ and $T_m = (T_t+T_b)/2$. Small corrections for the temperature 
gradients in the copper plates were applied. The center temperature $T_c$ was taken to 
be the average of the temperatures measured with eight thermistors attached to the 
outside of the side wall at the horizontal mid-plane, equally spaced in the azimuthal 
direction (see, for instance, Ref. \cite{bro06} or \cite{bro07b}). 

In  order to obtain an estimate of the OB 
values of the Nusselt numbers, a power law ${Nu}_{OB} = N_0 R^{\gamma_{eff}}$  was fitted to the Nusselt-number 
measurements at relatively small $\Delta$ where $\beta_m \Delta \alt 0.05$, adjusting $\gamma_{eff}$ and $N_0$. 
Such fits yielded values of $\gamma_{eff}$ close to 0.30. All the measured values of $Nu$ regardless of 
$\beta_m \Delta$ were then divided by the power-law value at the measured Rayleigh numbers to give ${Nu}/{Nu}_{OB}$ 
at all $\Delta$.

 All measurements reported here were made with many values of  $\Delta$ at each 
of a few constant values of $T_m$ and $P$. In both cells we used
ethane at elevated pressures as the fluid. The thermophysical properties were calculated 
from the formulas given in Ref.~\cite{fri91b}. For extensive discussions of the uncertainties of these properties we refer to that paper. It is difficult to determine the absolute errors for the Rayleigh and Nusselt numbers that results from property uncertainties, but we expect that an estimate of a few percent is not unreasonable. Since in the present paper we are concerned only with the ratios ${Nu}/{Nu}_{OB}$, and since all data are taken as a function of $\Delta$ at a given mean temperature and pressure and evaluated at the same $P$ and $T_m$, property errors cancel to a very large extent.

\subsection{Results}
\label{section/exp_results}

\subsubsection{The Nusselt number Nu(Ra)}

\begin{figure}
\begin{center}
\includegraphics*[width=8cm]{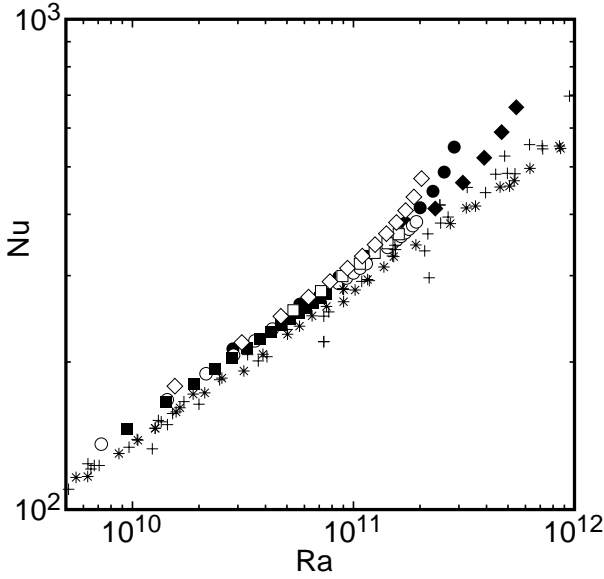}
\end{center}
\caption{(a): 
The Nusselt number Nu as a function of the Rayleigh number Ra for  $\Gamma = 1.00$. Solid circles: $P = 51.72$ bars and $T_m = 40^\circ$C (Pr =  2.58). Open circles: $P = 51.72$ bars and $T_m = 27^\circ$C (Pr = 2.99). Solid squares: $P = 51.72$ bars and $T_m = 24^\circ$C (Pr = 2.71).  Open squares: $P = 51.72$ bars and $T_m = 31^\circ$C (Pr = 3.85).  Solid diamonds: $P = 53.79$ bars and $T_m = 40^\circ$C (Pr = 3.79).   Open diamonds: $P = 50.00$ bars and $T_m = 40^\circ$C (Pr = 2.09). Stars: from Ref.~\cite{nie00} for $\Gamma = 0.5$ after correction for sidewall effects. Plusses: from Ref.~\cite{cha01} for $\Gamma = 0.5$.} 
\label{fig:nusselt}
\end{figure}

As indicated above, we do not regard the results for Nu(Ra) to be very accurate because of unknown effects due to the relatively large wall conductivity.  Nonetheless we show the results for $\Gamma = 1$ at several $T_m$ and $P$ in Fig.~\ref{fig:nusselt} on logarithmic scales. Over a wide range of Ra one sees that they are a few percent higher than the results from Refs.~\cite{nie00} and \cite{cha01}, and we attribute this to the influence of the side-wall conductivity on our data. At the largest Ra our results increase more rapidly with Ra, and data at different $T_m$ and $P$ begin to differ from each other. We attribute this phenomenon to NOB effects.

\begin{figure}
\begin{center}
\includegraphics*[width=8cm]{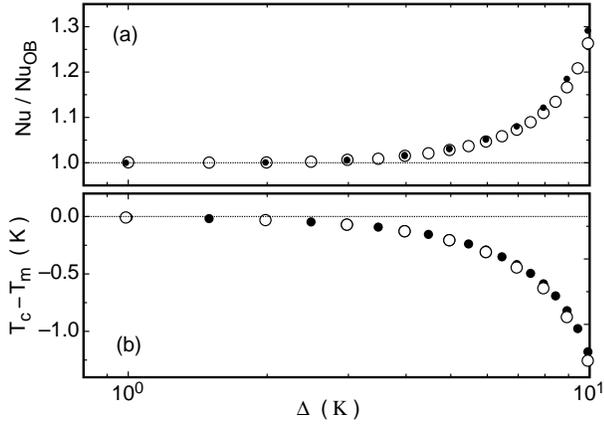}
\end{center}
\caption{(a): 
The ratio of the measured Nusselt number $Nu$ to the estimate ${Nu}_{OB}$ 
of the Nusselt number under Boussinesq conditions as a function of the applied temperature 
difference $\Delta$. (b) The deviation of the center temperature $T_c$ from the mean 
temperature $T_m$ as a function of $\Delta$. All measurements were made at $T_m = 40.00^\circ$C 
and a pressure of 51.72 bar ($P/P_*=1.062$) where the Prandtl number is 2.58. Open symbols: $\Gamma = 0.50$. 
Solid symbols: data  from Ref. \cite{ahl07} with $\Gamma = 1.00$.} 
\label{fig:Gamma_dependence}
\end{figure}

\subsubsection{Aspect-ratio dependence}
\label{sec:Gamma}
In Fig.~\ref{fig:Gamma_dependence} we compare results obtained at a mean temperature 
$T_m = 40.00^\circ$C and pressure $P = 51.72$ bar ($P/P_* = 1.062$, Prandtl number $Pr = 2.58$) in the 
sample of aspect ratio $\Gamma = 0.50$ (open circles) with previously reported results 
(\cite{ahl07}) for $\Gamma = 1.00$ (solid circles). One sees that the NOB effect on 
 ${Nu}$ and on the center temperature $T_c$ is within our resolution 
independent of $\Gamma$. This shows, as expected, that the NOB effects are confined 
essentially to the boundary layers. The length of the sample interior, which is 
nearly isothermal (see, however, Ref. \cite{bro07b}) regardless of its length, does not 
have a large influence. 
\begin{figure}
\begin{center}
\includegraphics*[width=8cm]{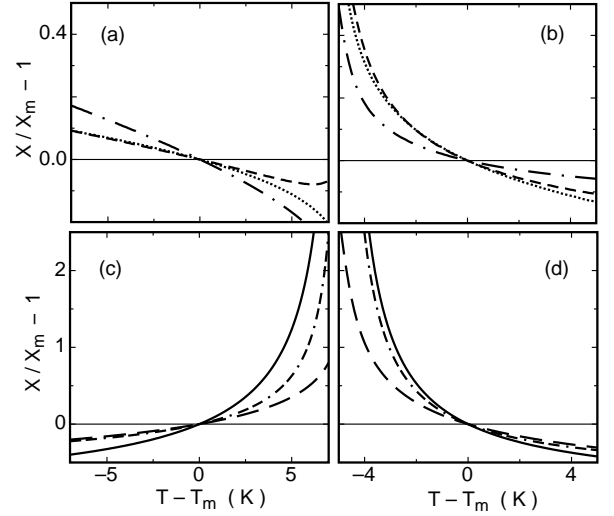}
\end{center}
\caption{The ratios $X/X_m$ for $P = 51.72$ bar ($P/P_*=1.062$) of several property values $X$ at 
temperatures $T-T_m$ to the value of $X$ at $T_m$  (based on Ref.~\cite{fri91b}). (a) and (c): $T_m = 27.00^\circ$C. 
(b) and (d): $T_m = 40.00^\circ$C. (a) and (b): thermal conductivity $\Lambda$ (short dashed line), 
density $\rho$ (dotted line), and dynamic viscosity $\eta$ (dash-dotted line). (c) and (d): 
thermal expansion coefficient $\beta$ (solid line), $\hat{\beta}$ (long dashed line), and heat capacity 
at constant pressure $c_P$ (double-dashed dotted line). 
The reference values $X_m$ for $27(40)^{\circ}$C are 
$\Lambda_m = 0.07328(0.04343)$~W m$^{-1}$ K$^{-1}$,
$\rho_m = 331.12(123.26)$~kg m$^{-3}$,
$\eta_m = 4.030(1.502)10^{-5}$~kg s$^{-1}$ m$^{-1}$,
$\beta_m = 0.01649(0.03815)$~K$^{-1}$, 
$c_{P,m}= 5434 (7452)$~J kg$^{-1}$ K$^{-1}$
and mean Prandtl number $Pr_m   =   2.99(2.58)$. 
} 
\label{fig:properties1}
\end{figure}
\begin{figure}
\begin{center}
\includegraphics*[width=8cm]{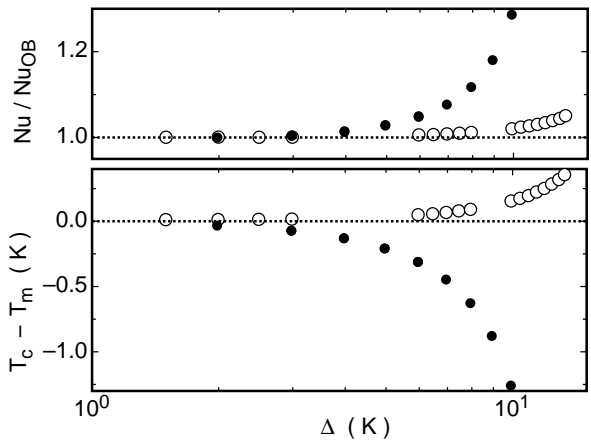}
\end{center}
\caption{(a): The ratio of the measured Nusselt number $Nu$ to the estimate ${Nu}_{OB}$ 
of the Nusselt number under Boussinesq conditions as a function of the applied temperature 
difference $\Delta$. (b) The deviation of the center temperature $T_c$ from the mean 
temperature $T_m$. These measurements were made for $\Gamma = 1.00$ at a pressure of 51.72 bar ($P/P_* = 1.062$). 
Open symbols: $T_m = 27.00^\circ$C  ($Pr = 2.99$). Solid symbols: $T_m = 40.00^\circ$C ($Pr = 2.58$).  These two 
temperatures are on opposite sides of, but not equidistant from,  the temperature  $T_\phi = 34.97^\circ$C where the 
critical isochore is reached on this isobar.} 
\label{fig:G=1.0}
\end{figure}

\subsubsection{Dependence on fluid properties}

Interesting insight into the influence of various property variations with temperature 
can be gained by measuring $T_c$ and $Nu$ along an isobar on the two sides of the 
temperature $T_\phi (P)$ at which the critical isochore is reached. In 
Fig.~\ref{fig:properties1} we show the variation along the isobar $P = 51.72$ bar $= 1.062 ~P_*$  
of the thermal conductivity $\Lambda$, density $\rho$, dynamic viscosity $\eta$, 
thermal expansion coefficient $\beta$, and heat capacity $c_P$ for the cases 
$T_m = 27.00^\circ$C (left panels) and 40.00$^\circ$C (right panels).
These two values are on opposite sides of but not quite equi-distant from $T_\phi = 34.97^\circ$C. 
In the upper two panels one sees that the 
variations of $\Lambda$, $\rho$, and $\eta$ are relatively small, have the same trends with $T$
though quantitatively they are somewhat different 
on the two sides, with maximum changes by less than a factor of two over temperature 
ranges that are small enough to avoid including $T_\phi$. On the other hand, the expansion 
coefficient and the heat capacity 
(lower two panels) vary by a factor of five or more. Thus one 
might expect them to dominate the NOB effects. Interestingly they have opposite trends with $T - T_m$,
the temperature derivatives of both $\beta$ and $c_P$ are positive below and negative above 
$T_\phi$; that is below the critical isochore (along the temperature axis), on the more liquid-like side,
$\beta$ and $c_P$ are smaller 
at the top (colder) than at the bottom (warmer) end of the sample, with this 
relationship reversed above the critical isochore, on the more gas-like side, where
$\beta$ and $c_P$ decrease from bottom to top.\\
In Fig.~\ref{fig:G=1.0} experimental results are presented for $\Gamma = 1.00$ at a pressure 
$P=51.72$ bar $= 1.062 ~P^*$. They are for the two mean temperatures $T_m = 27.00$ (open circles) 
and 40.00$^\circ$C (solid circles) of Fig.~\ref{fig:properties1} where the Prandtl numbers are 2.99 and 2.58 
respectively. In both cases we used $\Delta$ values sufficiently small so that $T_t$ ($T_b$) only reaches 
down (up) to $T_\phi$ so that the applied temperature 
difference does not straddle $T_\phi$. One sees that the NOB effects increase  $Nu$ on both sides of 
the critical isochore. On the high-temperature side (solid circles) the NOB effect is larger for the 
same $\Delta$. This is  consistent with the larger variation of the fluid properties at equal values 
of $T-T_m$ revealed above in Fig.~\ref{fig:properties1}.\\  
The NOB effect on $T_c$ 
is of {\it opposite} sign on the two sides of the critical isochore. For 
$T  <  T_\phi$ (open circles) NOB conditions increase $T_c$ above $T_m$, whereas 
for $T > T_\phi$ (solid circles) $T_c$ is reduced below $T_m$. This observation, 
in conjunction with the properties shown in Fig.~\ref{fig:properties1}, suggests 
that for these fluids the temperature drops $\Delta_{t,b}$ across the boundary 
layers are determined primarily by $\beta$ and/or $c_P$, with $\Delta_t > \Delta_b$ 
($\Delta_t < \Delta_b$) when $\beta$ and/or $c_P$ are smaller (larger) at the cold top 
end of the sample then at the warm bottom end. As pointed out before, \cite{ahl06} 
for the Nusselt number the contributions to the thermal resistance at the two boundary 
layers add, and it does not matter much whether the larger or smaller contribution 
comes from one end or the other. Thus, for $Nu$ the NOB effect is in the same direction 
in both cases. As was the case for $Nu$, the NOB effect revealed by $T_c$ is larger in magnitude 
above $T_\phi$ than it is  below. Again we attribute this difference  primarily to the difference 
in the variations of the properties shown in Fig.~\ref{fig:properties1}.
\begin{figure}
\begin{center}
\includegraphics*[width=8cm]{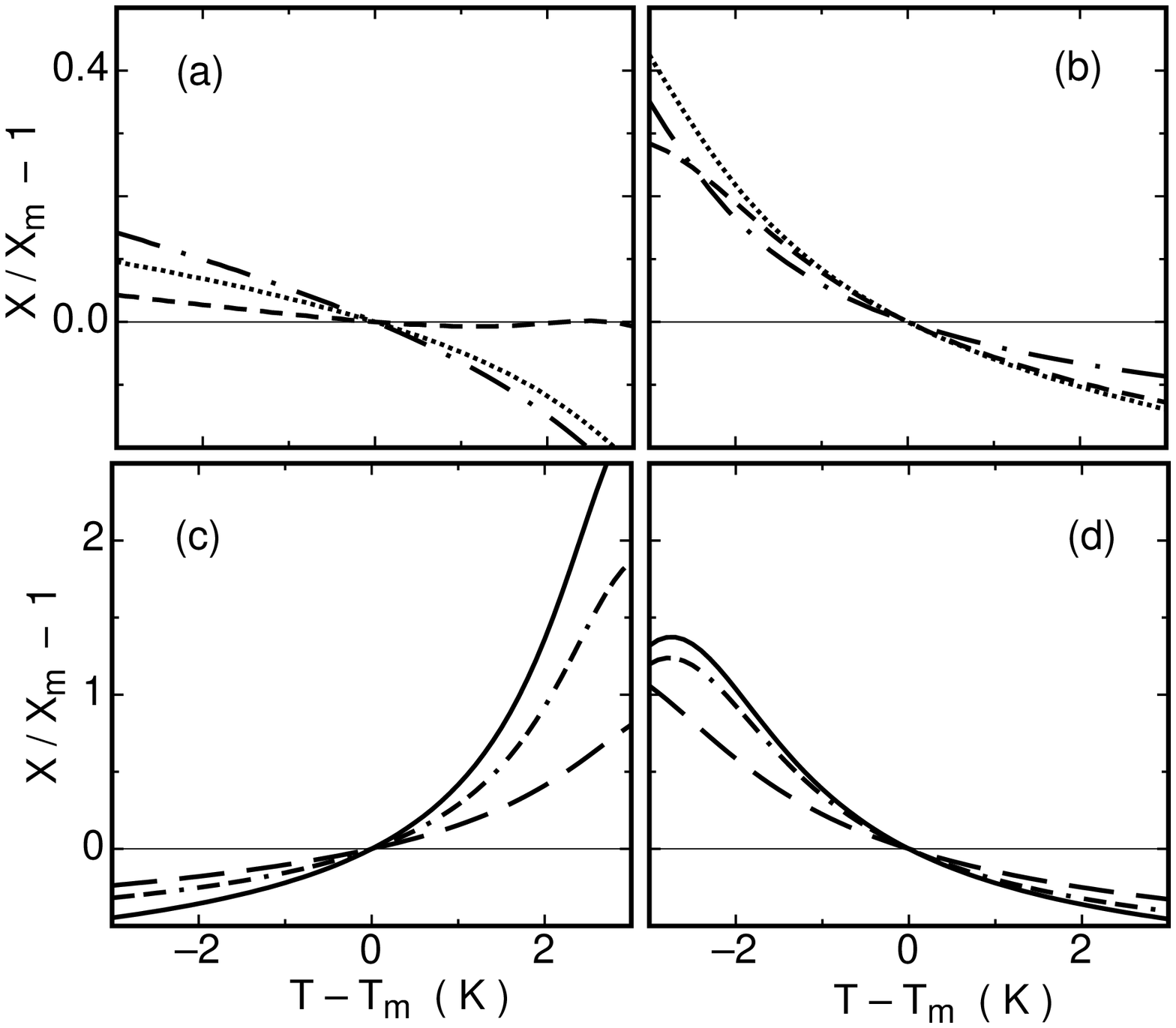}
\end{center}
\caption{The ratios $X/X_m$ for $P = 55.17$ bar ($P/P_* = 1.132$) of several property values $X$ at 
temperatures $T-T_m$ to the value of $X$ at $T_m$ (based on Ref.~\cite{fri91b}). (a) and (c): $T_m = 35.00^\circ$C. 
(b) and (d): $T_m = 41.00^\circ$C. (a) and (b): thermal conductivity $\Lambda$ (dashed line), 
density $\rho$ (dotted line), and dynamic viscosity $\eta$ (dash-dotted line). (c) and (d): 
thermal expansion coefficient $\beta$ (solid line), $\hat{\beta}$ 
(long dashed line), and heat 
capacity at constant pressure $c_P$ (double-dashed dotted line). The reference values $X_m$ for $35(41)^{\circ}$C are
$\Lambda_m = 0.06674(0.05163)$~W m$^{-1}$ K$^{-1}$, 
$\rho_m = 282.48(153.43)$~kg m$^{-3}$
$\eta_m = 3.168(1.726)10^{-5}$~kg s$^{-1}$ m$^{-1}$,
$\beta_m = 0.04177(0.06876)$~K$^{-1}$,
$c_{P,m}= 9617(12 534)$~J kg$^{-1}$ K$^{-1}$,
and mean Prandtl number $Pr_m   = 4.56 (4.20)$. 
} 
\label{fig:properties2}
\end{figure}

In  Fig.~\ref{fig:properties2} we show the variation along the isobar $P = 55.17$ bar 
($P/P_* = 1.132$) of the various properties for the cases 
$T_m = 35.00^\circ$C (left panels) and 41.00$^\circ$C (right panels). These 
two temperatures are also on opposite sides of and nearly equi-distant from the critical isochore,
for this pressure at $T_\phi = 38.06^\circ$C. 
Again the variation of the expansion coefficient and the heat capacity 
(lower two panels) is much larger than that of the other properties. 
At a given $|T-T_m|$, all the variations are more similar in  
magnitude on the two sides  of $T_\phi$ than they were for the case of Fig.~\ref{fig:properties1}.

\begin{figure}
\begin{center}
\includegraphics*[width=8cm]{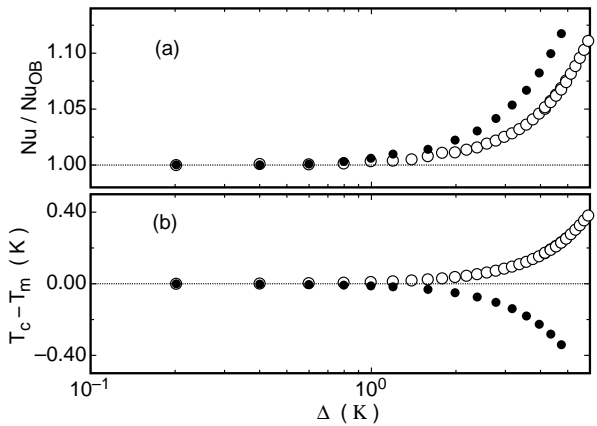}
\end{center}
\caption{(a): The ratio of the measured Nusselt number $Nu$ to the estimate ${Nu}_{OB}$ 
of the Nusselt number under Boussinesq conditions as a function of the applied temperature 
difference $\Delta$. (b) The deviation of the center temperature $T_c$ from the mean 
temperature $T_m$. These measurements were made for $\Gamma = 0.50$ at a pressure of 55.17 bar ($P/P_* = 1.132$). 
Open symbols: The mean temperature 
$T_m = 35^\circ$C  ($Pr = 4.56$). Solid symbols: $T_m = 41^\circ$C ($Pr = 4.20$).  These two temperatures are on 
opposite sides of and nearly equidistant from the temperature  $T_\phi = 38.06^\circ$C where the critical isochore 
is reached on this isobar.} 
\label{fig:G=0.5}
\end{figure}

In Fig.~\ref{fig:G=0.5} experimental results corresponding to the conditions of Fig.~\ref{fig:properties2} are 
presented for $\Gamma = 0.50$. They are for the two mean temperatures $T_m = 35.00$ (open circles) 
and 41.00$^\circ$C (solid circles) where the Prandtl numbers are 4.56 and 4.20 
respectively. In both cases we used $\Delta \leq 6$ K so that $T_t$ ($T_b$) reaches 
down (up) to $T_\phi$ when $T_m = 41^\circ$C ($35^\circ$C) while the applied temperature 
difference does not straddle $T_\phi$. For this case one sees that the NOB effects on $Nu$ are 
similar on the two sides of the critical isochore. Again, the NOB effect on $T_c$, 
although of about the same magnitude, is of opposite sign on the two sides. For 
$T  <  T_\phi$ (open circles, more liquid-like) NOB conditions increase $T_c$ above $T_m$, whereas 
for $T > T_\phi$ (solid circles, more gas-like) $T_c$ is reduced below $T_m$. 
\section{Comparison between boundary-layer theory and the experimental results}  
\label{section/comparison}

We now compare the experimental measurements from the previous section 
with the boundary layer results presented in section \ref{section/blt}. 
In particular, since the comparison with experiments at $T_m=40^{o}$ C (more gas-like ethane)
was already discussed in reference \cite{ahl07}, 
finding good agreement between experiment and the extended BL theory, we devote special 
attention to the new measurements at $T_m=27^{\circ}$ C and $P/P_{*} = 1.062$ (more liquid-like ethane).

\begin{figure}[!h]
\begin{center}
\epsfig{file=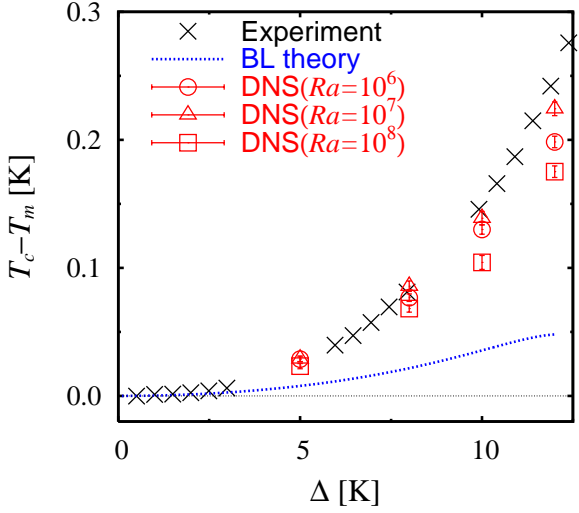,width=7.0cm,angle=-90}
\caption{
\label{figure/Tc-blexp} 
Deviation $T_c-T_m$ as function of $\Delta$, 
for $T_{m} = 27\,^{\circ}$C and $P/P_{*} = 1.062$. 
The symbols ($\times$) correspond to experimental 
measurements. The dotted line is obtained from boundary layer theory.
The red symbols 
($\bigcirc$, $\bigtriangleup$, and $\Box$) with error bars correspond 
to the incompressible DNS results described in section \ref{section/dns},
measured at different $Ra=10^6$ to $Ra = 10^8$.
Note that though these Rayleigh numbers are
smaller than in the experiments ($Ra=O(10^{9})$-$O(10^{10})$), 
the comparison is still appropriate because the $T_c$ shift has
proven to be rather independent 
of $Ra$ for given $\Delta$, provided one is  beyond the 
onset of the chaotic motion at $Ra\approx 2\cdot 10^5$ \cite{sug07,sug08}.
For further evidence for the weak $Ra$-dependence of 
the center temperature we  also refer to table \ref{tab_dns_tc_tm}.
}
\end{center}
\end{figure}

As shown in figure \ref{figure/Tc-blexp}, the curve for more liquid-like ethane obtained from BL theory 
considerably deviates from the experimental data. 
This is remarkably different from the comparison of BL theory with gaseous ethane,
presented in \cite{ahl07}, where instead a good agreement was observed.
This suggests that even though BL theory reasonably captures NOB effects associated with 
(OB.1) and (OB.2), further corrections are essential in the present liquid-like ethane case. 
Among them, deviations from (OB.3) seem to be the natural candidate for the failure of BL theory,
since the buoyancy force is \textit{not} included in the BL equations but apparently affects the 
thermal convection. Thus, 
in order to reveal the importance of nonlinear buoyancy in thermal convection, 
we shall perform direct numerical simulations (DNS) of the RB problem.

\section{Direct Numerical Simulations}
\label{section/dns}

\begin{figure}[!h]
  \begin{center}
\setlength{\unitlength}{1.0cm}
        \epsfig{file=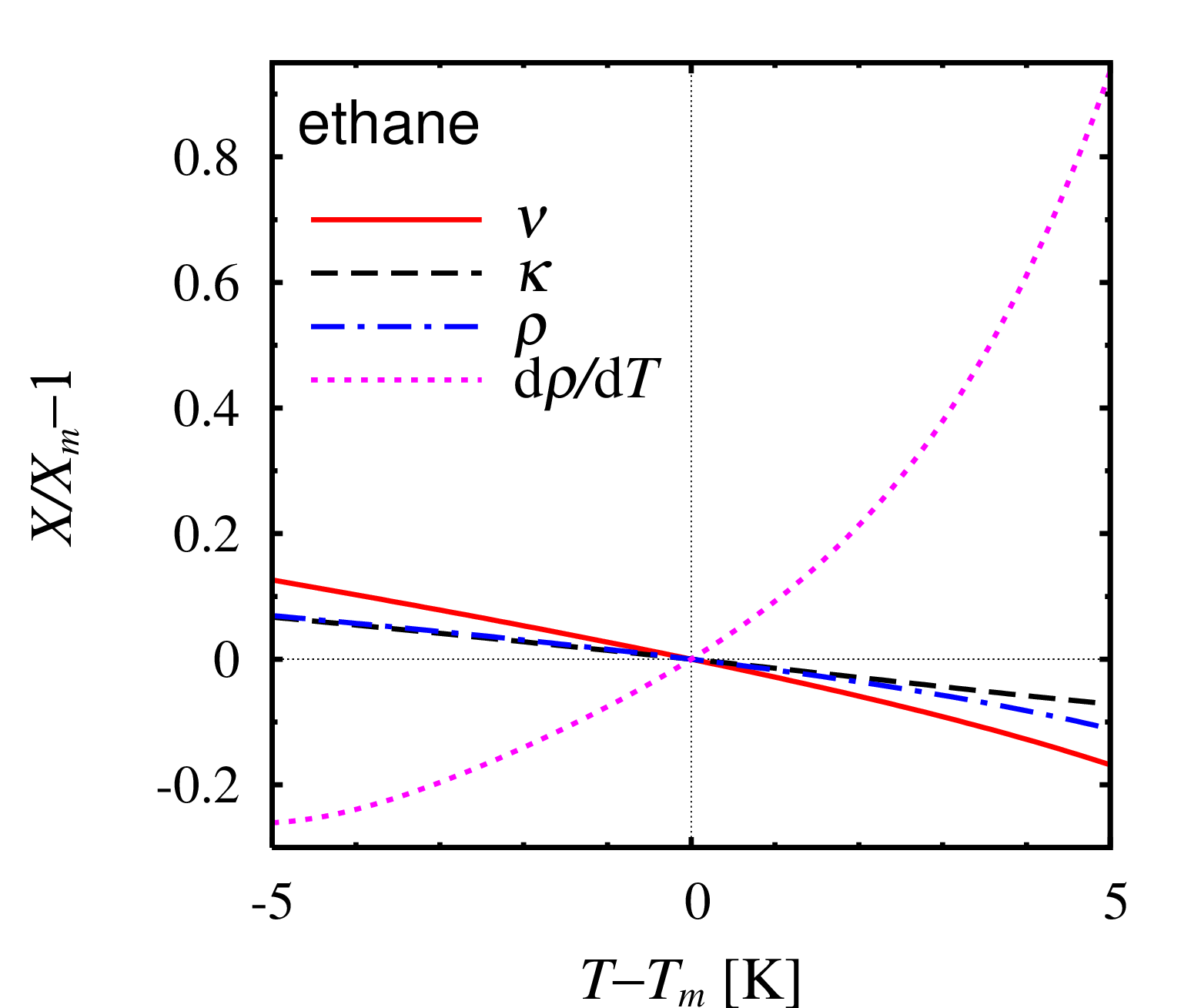,width=5.0cm,angle=-90}
        \epsfig{file=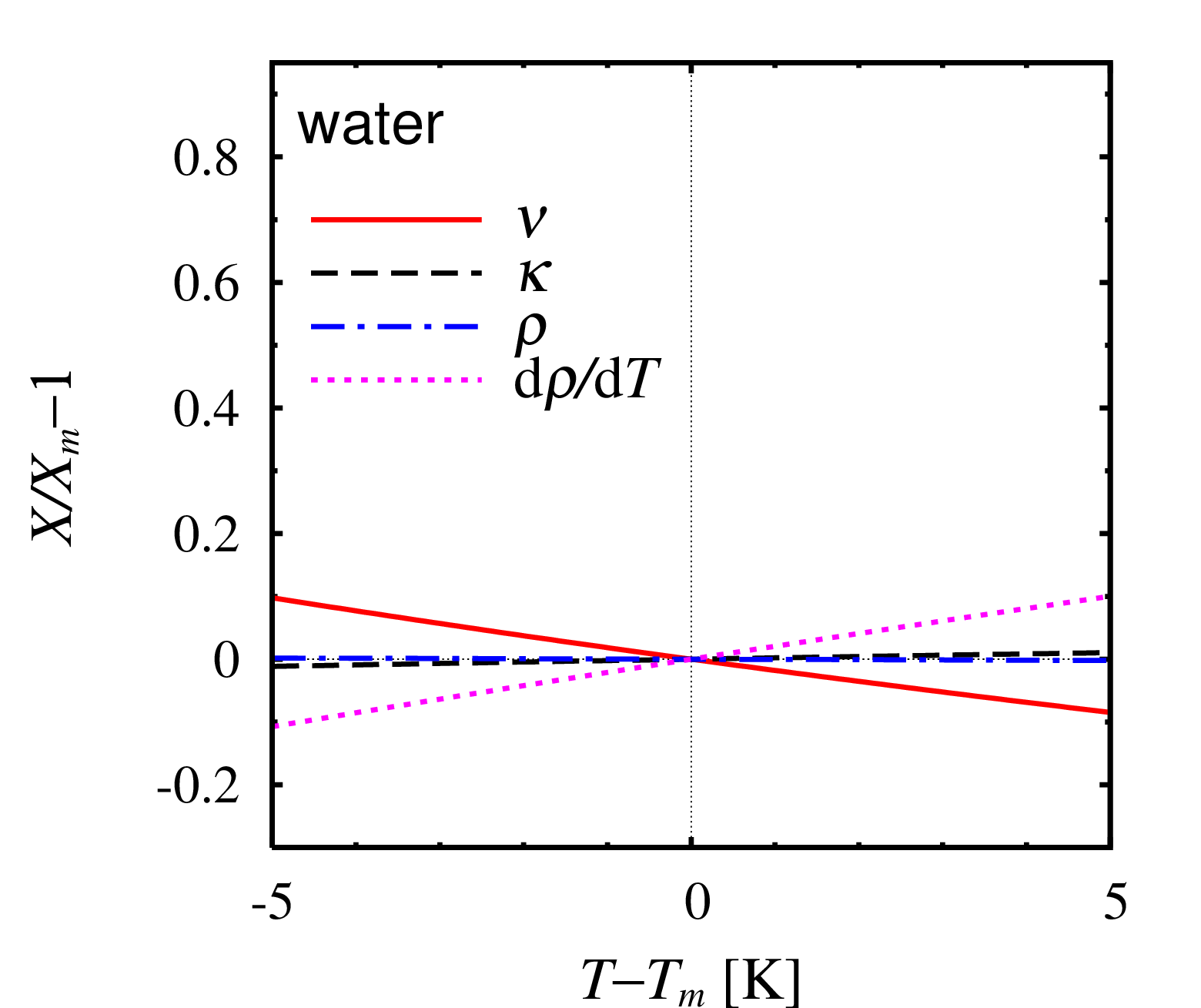 ,width=5.0cm,angle=-90}
        \epsfig{file=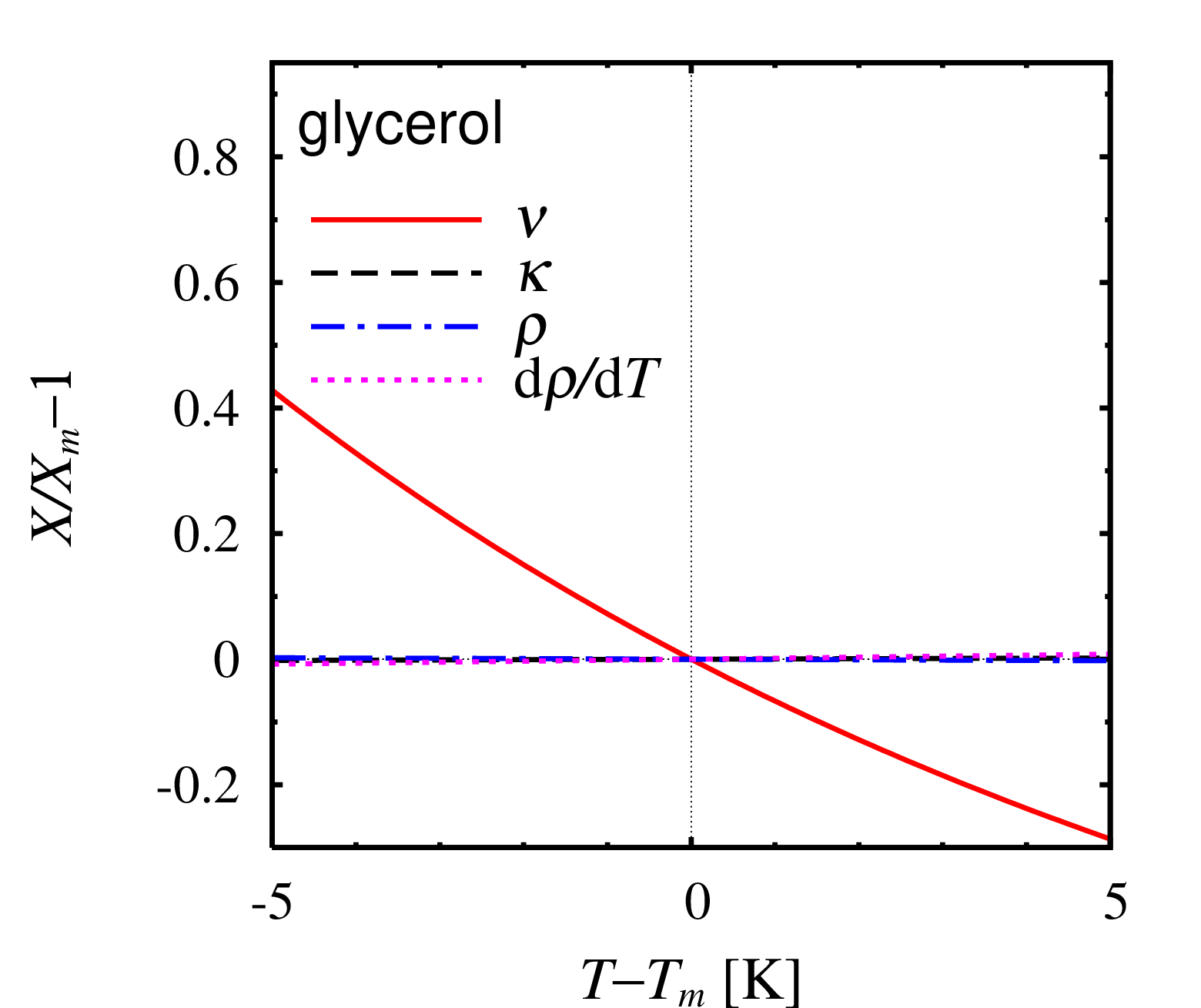,width=5.0cm,angle=-90}
  \end{center}
  \caption{Temperature dependences of the material properties $\nu$, $\kappa$,
    $\rho$ in the relevant $T$-range. ${\rm d}\rho/{\rm d}T$ is also displayed. (a) Ethane 
    around the temperature $T_m = 27^{\circ}$C adapted from \cite{fri91b}. 
    The pressure is $P/P_{*} = 1.062$. (b) and (c):
  The fluid properties for water (b) and glycerol (c), respectively, 
around the temperature 
    $T_m = 40^{\circ}$C (which has been studied in \cite{sug07} and \cite{sug08}). 
    Note the significant variation of the density $\rho$ with $T$ 
in the case of ethane, 
as compared to the two liquids.
In ethane, $\rho/\rho_m - 1$ various from about $0.07$ for $T-T_m = -5K$ 
to about $-0.1$ for $T-T_m = +5K$, whereas 
in glycerol $\rho\/rho_m$ is practically independent of $T$.
The strong variation of $\rho$ for ethane follows from the large 
    ${\rm d}\rho/{\rm d}T$ (also shown). 
}
  \label{fig_material_props}
\end{figure}

As shown in refs.\ \cite{sug07,sug08},
two-dimensional direct numerical simulations may be useful 
for the study of the tiny NOB effects which occur in RB convection 
in liquids. In particular, even with the restrictions to 
two-dimensional geometry and to incompressibility of the fluid flow, 
the effects on the center temperature and the Nusselt number shift could be reasonably 
captured in the cases of water and glycerol. 
However, the liquid-like ethane just above the critical pressure 
has a stronger temperature dependence 
of the density than water and glycerol. To quantify this, a comparison between the ethane properties 
around $T_m = 27^{\circ}$C and $P/P_{*} = 1.062$ with water and glycerol around 
$T_m = 40^{\circ}$C is reported in figure \ref{fig_material_props}. For the case of ethane the 
incompressible flow approximation seems to be questionable, or at least less justified. 
But we will show that adopting
the same approach used for water and glycerol \cite{sug07,sug08} also proves to be useful to study NOB 
effects in ethane and the results are well consistent with experiment.

Further insight into the problem is given by considering several cases of artificial 
ethane-like fluids, namely fluids which have only one, or some, of their material 
properties dependent on temperature, while the others are kept constant.  
In particular, as discussed in the previous section, we will examine the relevance of the 
nonlinear temperature dependence of buoyancy on the center temperature shift $T_c-T_m$ and 
take full notice of violating OB.3, which in contrast assumes constant $\partial \rho / \partial T$.
We remind that this cannot be taken into account in the extended BL theory 
presented in section \ref{section/blt}, while DNS can well include it. 

\subsection{Numerical simulation approach}

To handle the numerical effort we restrict ourselves to incompressible and even two dimensional flow.
The equations governing non-Oberbeck-Boussinesq convection in incompressible 
fluid flow are the incompressibility condition
\begin{equation}
\partial_i u_i=0 \ ,
\label{incompressibility}
\end{equation}
the Navier-Stokes equation
\begin{eqnarray}
\rho_m (\partial_t u_i + u_j \partial_j u_i) 
&=& 
-
\partial_i p  
+ 
\partial_j (\eta ( \partial_j u_i +\partial_i u_j )) \nonumber \\
&&
+ 
g \rho_m \left(1 - \rho/\rho_m \right)  \delta_{i3} ,
\label{ns}
\end{eqnarray}
and the heat-transfer equation
\begin{equation}
\rho_m c_{p,m} 
(\partial_t T + u_j \partial_j T ) = 
\partial_j ( \Lambda \partial_j  T ) .
\label{ht}
\end{equation}
Here, $\delta_{i3}$ is the Kroneker symbol.
The density  is assumed to be constant and its value $\rho_m$
is fixed at that of the temperature 
$T_m$, except in the buoyancy term, where the full nonlinear
temperature dependence of $\rho(T)$ is implemented. The dynamic viscosity $\eta(T)$ and the heat conductivity 
$\Lambda(T)$ are also both temperature and thus space dependent. The isobaric specific 
heat capacity $c_P$ is assumed to be constant, its value being $c_{p,m}$ (in contrast to real ethane). 
The experimentally known temperature dependences of $\rho$, $\eta$, $\Lambda$ and the values of the parameters 
$\rho_m$, $c_{p,m}$  for ethane are given in \cite{fri91b} and, for better reference, 
are reported in Table \ref{tab_material_prop} in the specific form implemented in our DNS.

For consistency with the experimental measurements 
and with the BL theoretical analyses presented above for liquid-like ethane, we chose the 
arithmetic mean temperature to be $T_m = 27^{\circ}$C and the pressure as $P/P_{*} = 1.062$.
\begin{table*}
{\small
\begin{tabular}{lrlrlrl}
\hline
&$\nu$&&$\kappa$&&$g(1-\rho/\rho_m)$&\\
$n$&$A_n$&&$B_n$&&$C_n$&\\\hline
0&$ 1.21734\cdot 10^{-7\ }$&[m$^2$/s]         &$ 4.07547\cdot 10^{-8\ }$&[m$^2$/s]	   &                      &                  \\
1&$-3.38861\cdot 10^{-9\ }$&[m$^2$/(s\ K)]    &$-5.77921\cdot 10^{-10 }$&[m$^2$/(s\ K)]    &$1.64833\cdot 10^{-2}$&[m/(s$^2$\ K)]    \\
2&$-8.30683\cdot 10^{-11 }$&[m$^2$/(s\ K$^2$)]&$-7.36395\cdot 10^{-12 }$&[m$^2$/(s\ K$^2$)]&$6.79967\cdot 10^{-4}$&[m/(s$^2$\ K$^2$)]\\
3&$-5.75280\cdot 10^{-12 }$&[m$^2$/(s\ K$^3$)]&$-9.06743\cdot 10^{-14 }$&[m$^2$/(s\ K$^3$)]&$4.53854\cdot 10^{-5}$&[m/(s$^2$\ K$^3$)]\\
4&$-7.64359\cdot 10^{-13 }$&[m$^2$/(s\ K$^4$)]&$ 1.49555\cdot 10^{-13 }$&[m$^2$/(s\ K$^4$)]&$6.13485\cdot 10^{-6}$&[m/(s$^2$\ K$^4$)]\\
5&$-8.70191\cdot 10^{-14 }$&[m$^2$/(s\ K$^5$)]&$ 2.56836\cdot 10^{-14 }$&[m$^2$/(s\ K$^5$)]&$6.94645\cdot 10^{-7}$&[m/(s$^2$\ K$^5$)]\\
\hline
\end{tabular}
}
\caption{ 
Expansion coefficients of material properties of ethane around the temperature 
$T_m = 27^{\circ}$C adapted from \cite{fri91b}.
The pressure normalized by its critical value is $P/P_{*} = 1.062$.
The {\it effective} kinematic viscosity, 
the {\it effective} thermal diffusivity, and the buoyancy
are written in a polynomial form as 
$\nu(T) \equiv \eta(T)/\rho_m$ $=$ 
$\sum_{n=0}A_n (T-T_m)^n$ $[{\rm m}^2/{\rm s}]$,
$\kappa(T) \equiv \Lambda(T)/(\rho_m c_{p,m})$ 
$=$ $\sum_{n=0}B_n (T-T_m)^n$ $[{\rm m}^2/{\rm s}]$,
and $g(1-\rho(T)/\rho_m)$$=$$\sum_{n=1}C_n (T-T_m)^n$ $[{\rm m}/{\rm s}^2]$, 
respectively. Using the leading coefficient for the buoyancy force,
we can write the Rayleigh number as $Ra=\beta_m L^3\Delta/(\nu_m \kappa_m)$, 
where $\beta_m=C_1$, $\nu_m=A_0$ and $\kappa_m=B_0$, which coincides with the usual OB 
definition. The polynomial expensions for $\beta(T)$ and $\hat{\beta}(T)$ are $g \hat{\beta}(T) = 
\sum_{n=1} C_n (T-T_m)^{n-1}$ and $g \beta (T) = (\rho_m / \rho(T)) \cdot \sum_{n=1} n C_n (T-T_m)^{n-1}$.
}
\label{tab_material_prop}
\end{table*}
\subsection{Numerical results: $T_c$ shift in liquid-like ethane}
From figure \ref{figure/Tc-blexp} we can conclude that the DNS captures the experimental 
measurements of the center temperature shift $T_c-T_m$ as a function of $\Delta$ quite reasonably.
The quality of the agreement with the available ethane data is 
similar to the one we have observed for glycerol \cite{sug07} and for water \cite{sug08}. 
This also serves as a further validation of our numerical approach. 

We note that for water and for glycerol the $T_c$ shift obtained by the extended BL theory \cite{ahl06} 
is nearly the same as calculated by DNS (see \cite{sug07,sug08}). In contrast,
for the liquid-like ethane, the extended and even compressible BL theory only provides 
the right trend in the shift, but can not capture its amplitude (see again Fig. \ref{figure/Tc-blexp}).
This observation supports our guess on the relevance of the nonlinear $T$-dependence of $\rho(T)$ and thus 
of buoyancy. This latter is fully included in the DNS, as described in the previous subsection, 
while in BL theory it cannot be taken care of. 

One of the advantages of the DNS as compared to real experiments is that the material properties are 
easily and independently tunable. Therefore, the dynamics of \textit{hypothetical ethane-like liquids} 
can also be addressed. In the next subsection we shall see how this approach is useful in 
understanding the effects of the temperature dependence of the various material properties 
on the center temperature shift.

\begin{figure}
\epsfig{file=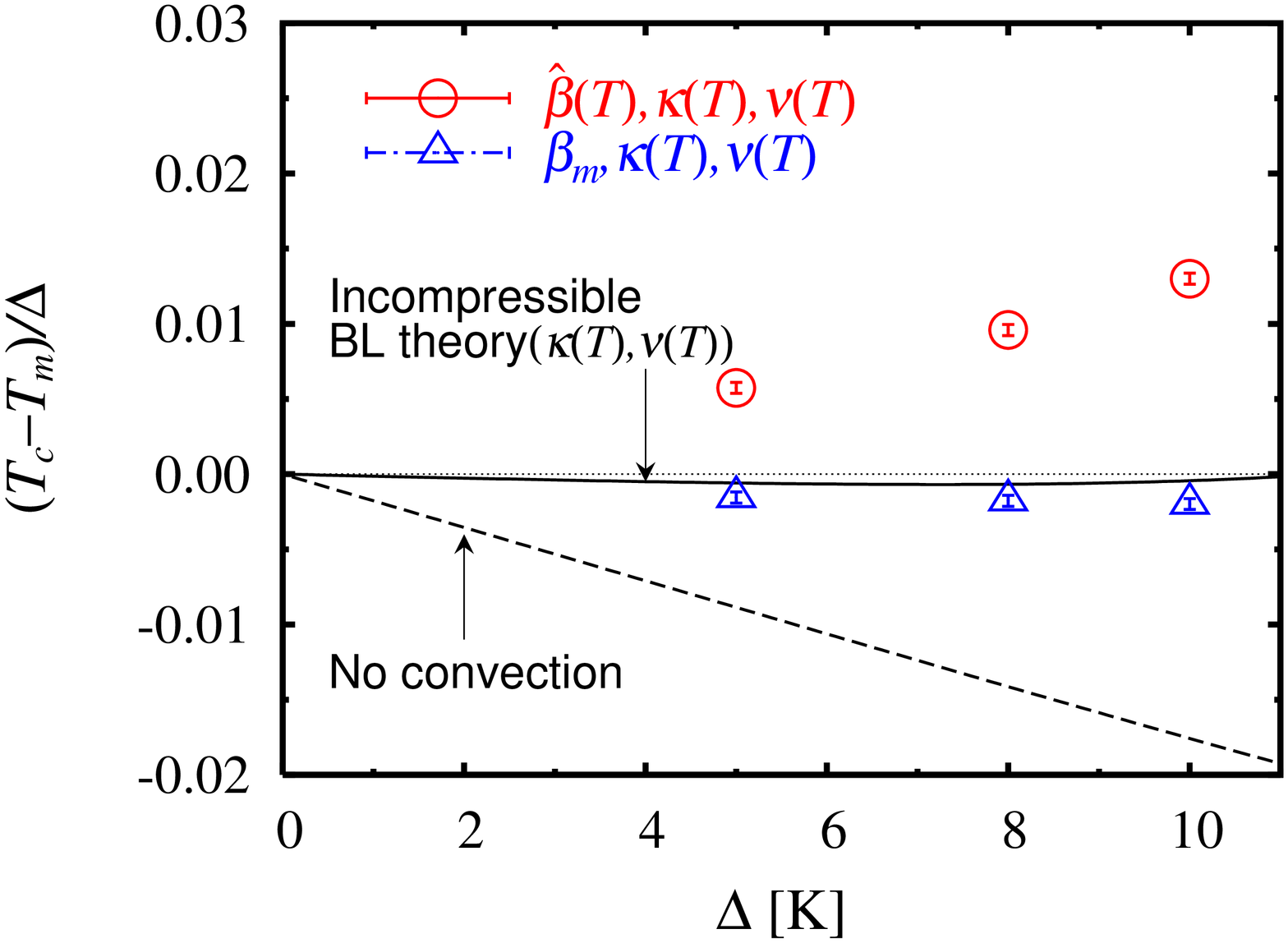,width=6cm,angle=-90}
\epsfig{file=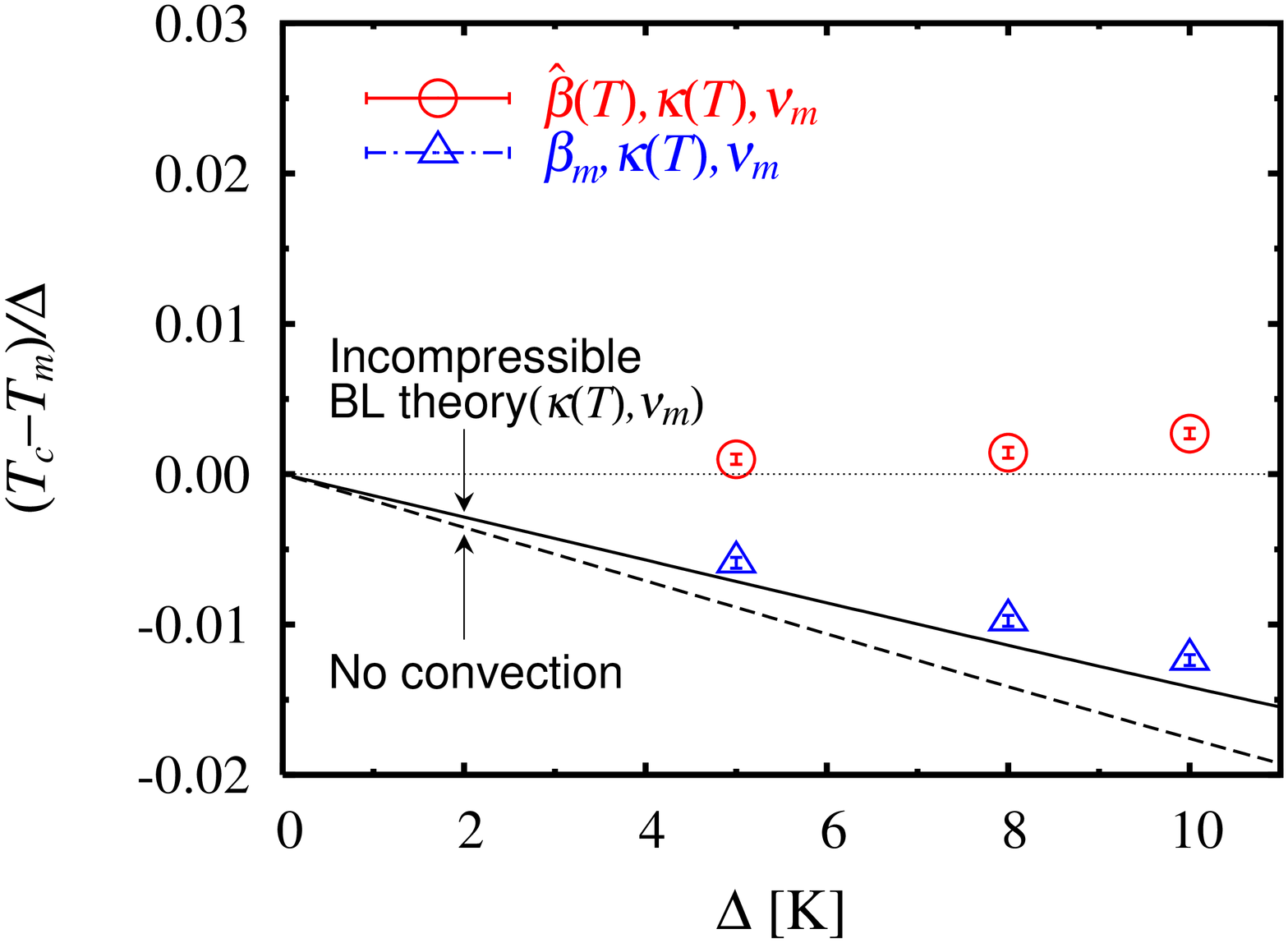 ,width=6cm,angle=-90}
\epsfig{file=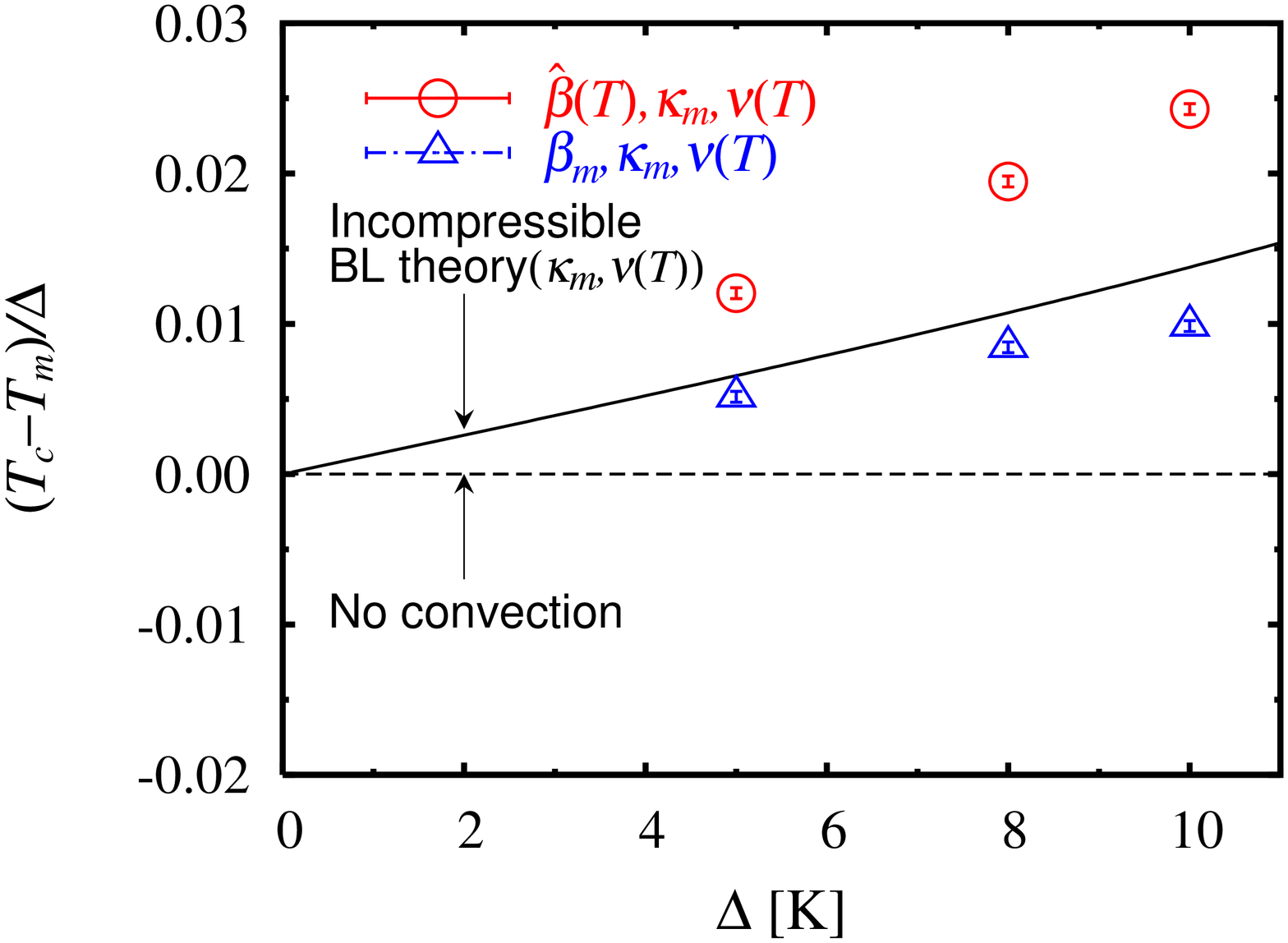 ,width=6cm,angle=-90}
\caption{
The normalized center temperature shift $(T_c-T_m)/\Delta$ versus 
the temperature difference $\Delta$ for several hypothetical liquids.
We consider the following six hypothetical liquids:  
$(\hat{\beta}(T), \kappa(T), \nu(T))$, $(\beta_m, \kappa(T), \nu(T))$,
$(\hat{\beta}(T), \kappa(T), \nu_m)$, $(\beta_m, \kappa(T), \nu_m)$, 
$(\hat{\beta}(T),\kappa_m, \nu(T))$ and $(\beta_m, \kappa_m, \nu(T))$.
In particular in each panel we compare two cases, which differ only by their buoyancy's $T$-dependence, i.e., 
$\hat{\beta}(T)$ instead of $\beta_m$.  
The symbols indicate the simulation results at Rayleigh number $Ra=10^6$. The circles ($\bigcirc$) 
correspond to the cases in which the full temperature dependence of the buoyancy $\left(T-T_m \right) 
\hat{\beta}(T)$ is taken into account, while the triangles ($\triangle$) represent the cases where only the linear 
temperature dependence $\left( T - T_m \right) \beta_m $ is considered. 
The solid line shows the prediction of boundary layer theory with an incompressible 
flow assumption \cite{ahl06}. The dashed line stems from the solution with no convective 
flow $\partial_z ( \kappa(T) \partial_z T )=0$ 
with boundary conditions $T=T_b$ at $z=0$ and $T=T_t$ at $z=L$.
See also Table \ref{tab_dns_tc_tm}.
}
\label{fig_dns_compare_w_blt}
\end{figure}

\subsection{$T_c$ shift in hypothetical fluids}

To obtain more insight into the physical origin of the non-Oberbeck-Boussinesq $T_c$-shift,
we consider NOB corrections for hypothetical ethane-like fluids
in which at least one of the temperature dependences of $\kappa(T)$, $\nu(T)$,
$\hat{\beta}(T)$ 
is switched off, and fixed at the OB values $\kappa_m$, $\nu_m$, $\beta_m$.
The quantity $\hat{\beta}(T)$, defined in eq.\ (\ref{hatbeta}), 
is useful for the classification of the hypothetical fluids discussed 
in the following sections, but $\hat{\beta}(T)$ is not explicitly introduced into the DNS, 
in which the density difference $\rho(T) - \rho_m$ is taken instead, 
see Eq.(\ref{ns}) 
and Tab. \ref{tab_material_prop}. Finally, we remind that $\hat{\beta}(T) = \beta_m$, the usual 
thermal expansion coefficient, if the fluid density is a \textit
{linear} function of the temperature around $T_m$ (i.e., if the conditions OB.3 holds).
In that case the thermal expansion coefficient $\beta(T) = \beta_m / [1 - (T-T_m) \beta_m]$ 
still depends on $T$ unless $\beta \Delta \ll 1$.  

For convenience hereafter we will call the two classes of artificial 
fluids, based respectively on the full non-linear NOB buoyancy force and on the linear OB approximation
as defined by OB.3, as the $\hat{\beta}(T)$-fluids and $\beta_m$-fluids.

\begin{table*}

\begin{center}
{\small
\begin{tabular}{llllrrr}
\hline
case&$\hat{\beta}$&$\kappa$&$\nu$&$100(T_c-T_m)/\Delta$&$100(T_c-T_m)/\Delta$&$100(T_c-T_m)/\Delta$\\
&&&&at $Ra=10^6$&at $Ra=10^7$&at $Ra=10^8$\\\hline
1 (NOB)&$\hat{\beta}(T)$     &$\kappa(T)$&$\nu(T)$&$ 1.3003\pm 0.0369$&$ 1.3879\pm 0.0527$&$ 1.0788\pm 0.0362$\\
2      &$\hat{\beta}(T)$     &$\kappa(T)$&$\nu_m$ &$ 0.2699\pm 0.0356$&$ 0.2174\pm 0.0515$&$ 0.1251\pm 0.0414$\\
3      &$\hat{\beta}(T)$     &$\kappa_m$ &$\nu(T)$&$ 2.4283\pm 0.0361$&$ 2.4351\pm 0.0496$&$ 2.4370\pm 0.0462$\\
4      &$\hat{\beta}(T)$     &$\kappa_m$ &$\nu_m$ &$ 1.4805\pm 0.0355$&$ 1.4320\pm 0.0735$&$ 1.3796\pm 0.0534$\\
5      &$\hat{\beta}(2T_m-T)$&$\kappa(T)$&$\nu(T)$&$-1.5953\pm 0.0361$&$-1.7946\pm 0.0506$&$-1.5868\pm 0.0320$\\
6      &$\hat{\beta}(2T_m-T)$&$\kappa(T)$&$\nu_m$ &$-2.6382\pm 0.0357$&$-2.7458\pm 0.0496$&$-2.5516\pm 0.0496$\\
7      &$\hat{\beta}(2T_m-T)$&$\kappa_m$ &$\nu(T)$&$-0.4738\pm 0.0375$&$-0.4578\pm 0.0580$&$-0.3358\pm 0.0354$\\
8      &$\hat{\beta}(2T_m-T)$&$\kappa_m$ &$\nu_m$ &$-1.4878\pm 0.0366$&$-1.5358\pm 0.0504$&$-1.4013\pm 0.0363$\\
9      &$\beta_m$            &$\kappa(T)$&$\nu(T)$&$-0.1983\pm 0.0361$&$-0.2043\pm 0.0467$&$-0.2691\pm 0.0587$\\
10     &$\beta_m$            &$\kappa(T)$&$\nu_m$ &$-1.2369\pm 0.0363$&$-1.1883\pm 0.0611$&$-1.1851\pm 0.0369$\\
11     &$\beta_m$            &$\kappa_m$ &$\nu(T)$&$ 0.9852\pm 0.0364$&$ 1.0834\pm 0.0576$&$ 0.9916\pm 0.0259$\\
12 (OB)&$\beta_m$            &$\kappa_m$ &$\nu_m$ &$ 0.0171\pm 0.0381$&$-0.0484\pm 0.0548$&$ 0.0271\pm 0.0387$\\
\hline
\end{tabular}
}
\end{center}

\caption{ 
DNS results for the center temperature shift $T_c-T_m$ 
normalized by the temperature difference $\Delta=10$K 
for several hypothetical fluids. 
The {\it effective} thermal expansion function is given by 
$\hat{\beta}(T)$ $=$ 
$g^{-1}\sum_{n=1} C_n (T-T_m)^{n-1}$ $[1/{\rm K}]$. 
Using the expansion coefficients $C_n$ listed in Table \ref{tab_material_prop},
we write the buoyancy for the case of $\hat{\beta}(2T-T_m)$ as 
$g(1-\rho(T)/\rho_m)$$=$$\sum_{n=1}(-1)^{n+1}C_n (T-T_m)^n$ $[{\rm m}/{\rm s}^2]$,
and in the $\beta_m$ case as $g(1-\rho(T)/\rho_m)$$=$$C_1(T-T_m)$ $[{\rm m}/{\rm s}^2]$.
Although the center temperature shift for the OB case (case 12) should be essentially zero due 
to the top-bottom symmetry,
the mean value determined from the DNS result is non-zero since the sampling time for taking the 
statistics is finite. 
Note that measurements of the temperature shift in the OB case are all, within statistical uncertainty, compatible with zero.
}
\label{tab_dns_tc_tm}
\end{table*}

In figure \ref{fig_dns_compare_w_blt} we present the DNS results of the normalized temperature shift 
$(T_c-T_m)/\Delta$ for several types of hypothetical fluids. 
The numeral values are given in table \ref{tab_dns_tc_tm}.
One clearly observes in the figures and in the table that the temperature
dependence of the thermal expansion function $\hat{\beta}(T)$ apparently is 
relevant for the shift of $T_c$. 
What can also be noticed in particular from table \ref{tab_dns_tc_tm} is that
the corrections of the center temperature originating from 
the temperature dependence of either $\nu$, $\kappa$, or $\beta$ are 
approximately additive (i.e., add "linearly"): E.g., the center temperature
corrections of the fluids with ($\beta_m$, $\kappa_m$, $\nu (T)$) and with 
($\beta_m$, $\kappa (T)$, $\nu_m$) add to that of the fluid with 
($\beta_m$, $\kappa (T)$, $\nu (T)$), etc. Note that this additivity
is in contrast to what had been found within the extended BL theory
of ref.\ \cite{sug07} where the full compressibility of the density
had been taken into consideration. Instead, in DNS we have restricted to incompressible flow.

\subsubsection{$\hat{\beta}(T)$- and $\beta_m$-fluids and extended BL}
In figure \ref{fig_dns_compare_w_blt} we also compare 
fluids of $\hat{\beta}(T)$- and $\beta_m$-type with equal properties of thermal conductivity 
and kinematic viscosity on the different panels. 
For convenient comparison the predictions by the extended BL theory under the assumption of 
fluid incompressibility \cite {ahl06} are also shown for the hypothetical fluids. 
Note that in such case the chosen type of buoyancy force needs not be specified, because BL theory does not 
capture it: For the BL theory $\hat{\beta}(T)$- and $\beta_m$-ethane are not distinguishable with respect to buoyancy.
As a reference, the $T_c$-shift value in the purely conductive case ($u_i = 0$ everywhere) is also reported. 
The $T_c$ shift in this case is not zero for fluids with $\kappa=\kappa(T)$ as a result of the solution of 
the heat conduction equation $\kappa\ \partial_z^2 T  +\frac{{\rm d}\kappa}{{\rm d}T}( \partial_z T)^2 =0$
with $\frac{\partial \kappa}{\partial T} |_P \neq 0$. 

We now discuss our main findings.
The change in $T_c$ for the hypothetical liquids with $\beta_m$ relative to the ones with $\hat{\beta}(T)$ 
is comparable for given $\Delta$.
More importantly, we find that the deviation $(T_c-T_m)/\Delta$ calculated with DNS is well captured 
by the BL theory for the cases of artificial $\beta_m$-fluids, i.e., for fluids where $\rho(T)$ is 
assumed to be a linear function of $T$. In contrast, BL theory is always far from the $T_c$ shift 
obtained for the more real $\hat{\beta}(T)$-ethane.
This indicates that the extended BL theory well captures the NOB effect once
the assumption (OB.3) is satisfied, even if (OB.1) and (OB.2) are violated,
but it does {\it not} correctly describe the NOB effects, if (OB.3) is violated.
Furthermore, the DNS results reveal that the $T_c$-shift is always enhanced if a temperature dependence of 
the thermal expansion function $\hat{\beta}(T)$ determines the buoyancy, i.e., if $\rho(T)$ 
depends nonlinearly on $T$.

\subsubsection{Mirror transformation $\hat{\beta}(T) \to \hat{\beta}(2T_m - T)$}
To quantitatively appreciate the effect of the temperature dependence in each material property individually
also at different Rayleigh numbers, $Ra=10^6 $-$10^8$, we list the $T_c$ shifts for several hypothetical 
ethane-like fluids in Tab. \ref{tab_dns_tc_tm}. 
Since here our primary concern is the influence of the thermal expansion function, 
besides the $\hat{\beta}(T)$- and $\beta_m$-fluids a new class of hypothetical fluids is introduced. 
We consider $\hat{\beta}(2T_m-T)$-fluids obtained by the mirror transformation 
$\hat{\beta}(T) \to \hat{\beta}(2T_m - T)$.
\begin{figure}
\begin{center}
\epsfig{file=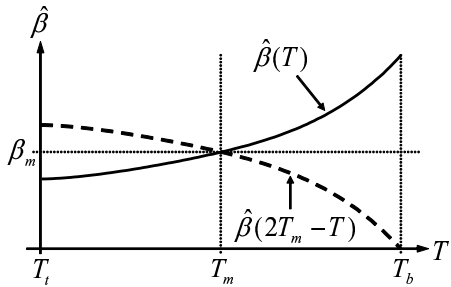,width=6cm,angle=0}
\end{center}
\caption{
Schematic plot of the mirror transformation $\hat{\beta}(T) \rightarrow \hat{\beta}(2T_m-T)$
of the thermal expansion coefficient $\hat{\beta}$.
}
\label{fig_schem_hat_bet}
\end{figure}
As schematically shown in figure \ref{fig_schem_hat_bet}, 
this transformation reverts the nonlinearity in the buoyancy force with respect to $T-T_m$.
The comparison between the cases $(\hat{\beta}(T), \kappa_m, \nu_m)$ and 
$(\hat{\beta}(2T_m -T), \kappa_m, \nu_m)$ - cases 4 and 8 in table \ref{tab_dns_tc_tm} - 
shows that the effect of the mirror transformation on the output parameter $T_c$
is to change the sign of $(T_c-T_m)/\Delta$ while preserving its modulus.
Furthermore, the deviation of $(T_c-T_m)/\Delta$ for all $\beta_m$-fluids (cases 9,10,11) relative
to the $\hat{\beta}(T)$-fluids (cases 1,2,3) is always positive, while it is always negative relative to the 
$\hat{\beta}(2T_m-T)$-fluids (cases 5,6,7). 
These features hold at all the studied $Ra$ numbers. Therefore, we conclude that the shift $T_c-T_m$ 
is sensitive to the sign of the slope of $\hat{\beta}(T)$ or, equivalently, to
the sign of the nonlinear term in the buoyancy factor $1-\rho(T)/\rho_m$. 
More precisely speaking, the mirror transformation changes 
the signs of the even order coefficients $C_2$ and $C_4$ defined in Tab.\ \ref{tab_material_prop}.
Obviously $C_2$ is the larger and thus the most relevant coefficient.
These features are absent in water and in glycerol, because for those the temperature 
dependence of $\rho(T)$ is much less pronounced. 
\begin{figure}
\begin{center}
\epsfig{file=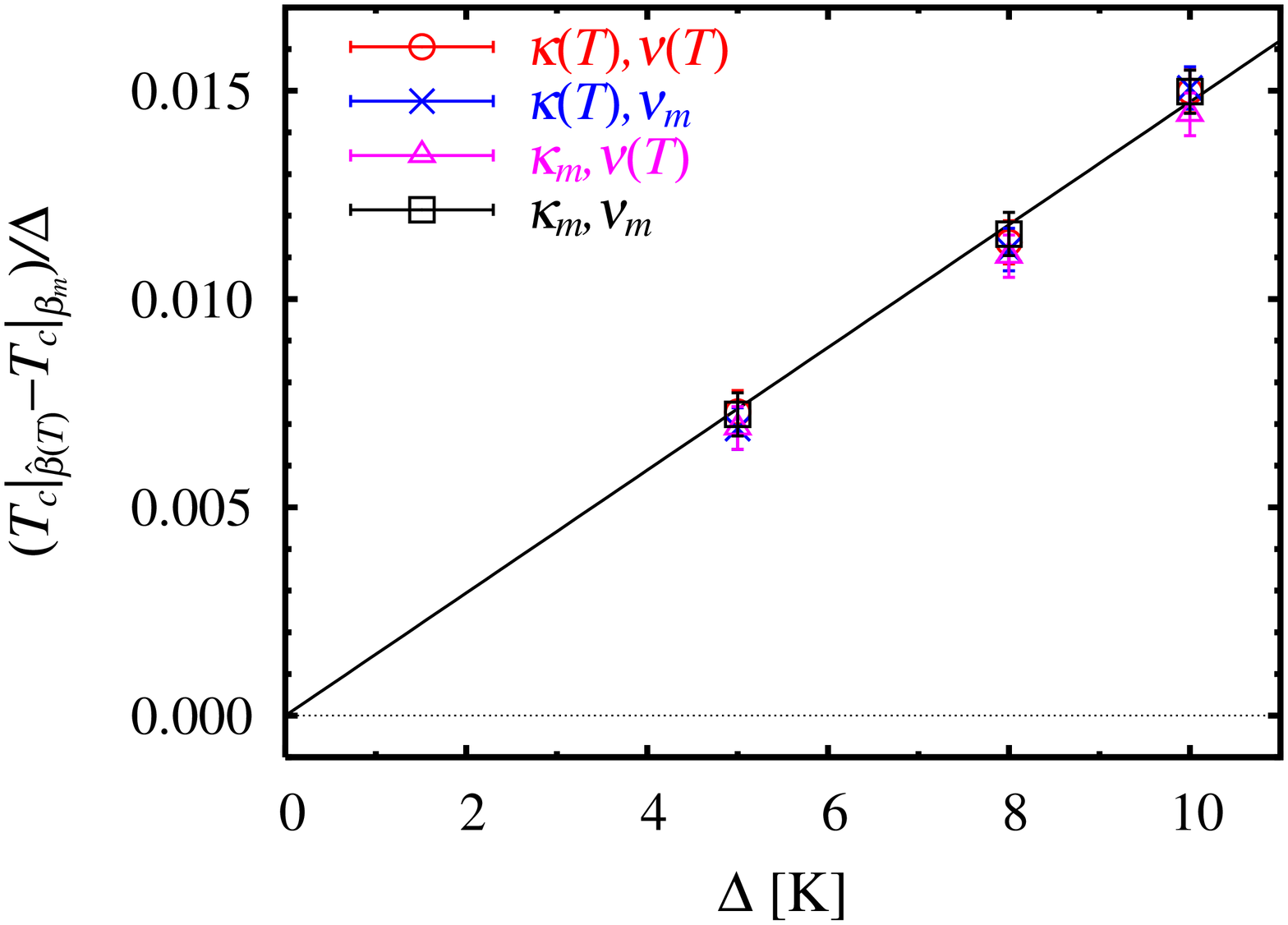,width=6cm,angle=-90}
\epsfig{file=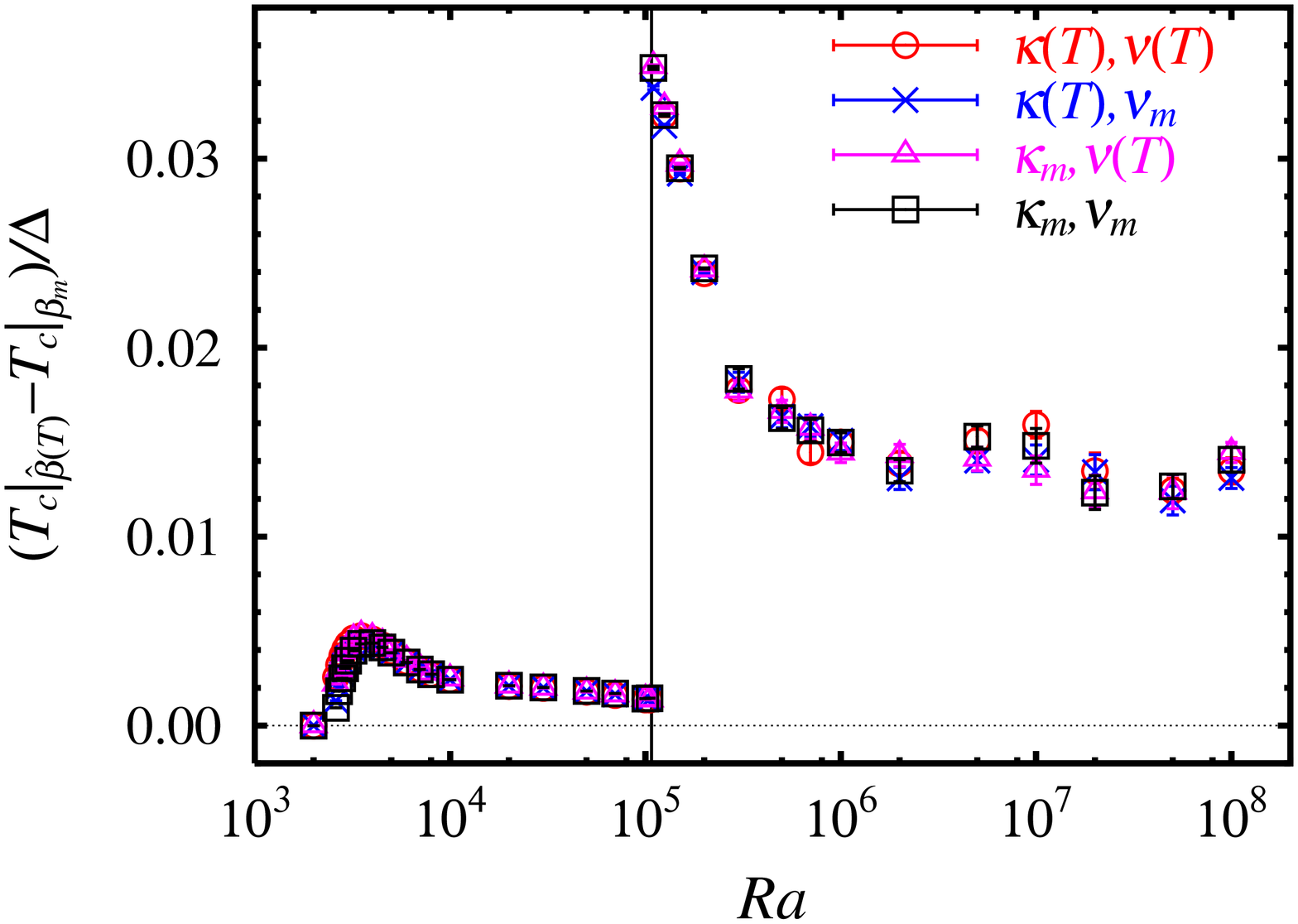 ,width=6cm,angle=-90}
\epsfig{file=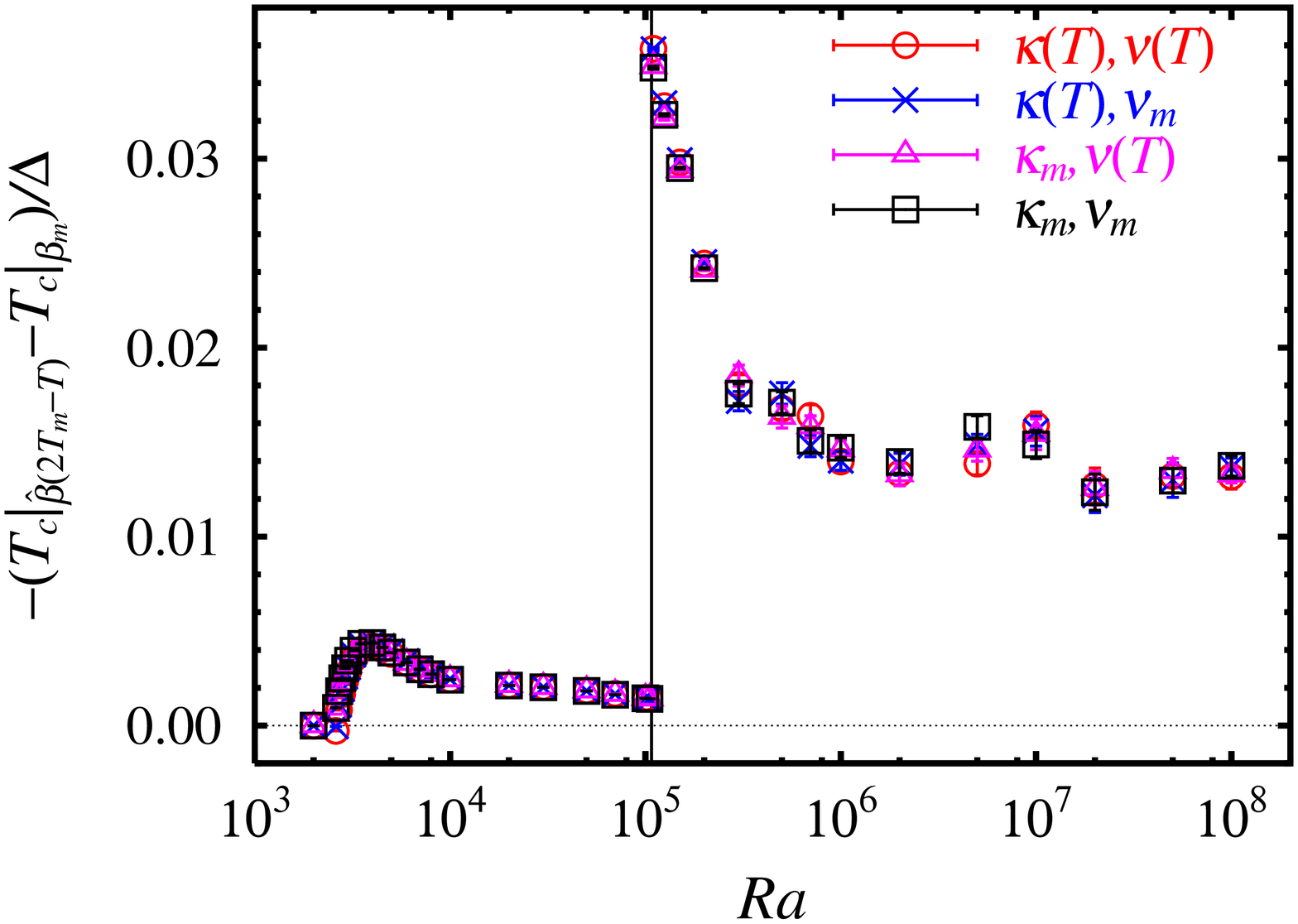,width=6cm,angle=-90}
\end{center}
\caption{
Effect of the temperature dependence of the thermal expansion coefficient $\hat{\beta}$ on 
the shift of the center temperature for several hypothetical fluids. DNS 
results for the normalized temperature difference $(T_c|_{\hat{\beta}(T)}-T_c|_{\beta_m})/\Delta$
(or  $-(T_c|_{\hat{\beta}(2T_m-T)}-T_c|_{\beta_m})/\Delta$)
with the same temperature dependence of the thermal diffusivity $\kappa$ and the kinematic 
viscosity $\nu$ are plotted; here $T_c|_{\hat{\beta}(T)}$ denotes the center temperature with 
the full temperature dependence of the buoyancy $g(1-\rho/\rho_m)$ as given in 
Table \ref{tab_material_prop}, and $T_c|_{\beta_m}$ denotes that with the linear 
temperature dependence $g(1-\rho(T)/\rho_m)$$=$$C_1(T-T_m)$ only. The top panel shows 
$(T_c|_{\hat{\beta}(T)}-T_c|_{\beta_m})/\Delta$ versus $\Delta$ at fixed Rayleigh number 
$Ra=10^6$. The solid line shows the linear fit
$1.473\cdot 10^{-3}{\rm K^{-1}}\times \Delta$ for the case 
$\hat{\beta}(T), \kappa_m, \nu_m$ (see Table \ref{tab_dns_tc_tm}).
The middle panel shows $(T_c|_{\hat{\beta}(T)}-T_c|_{\beta_m})/\Delta$ versus $Ra$ at 
fixed temperature difference $\Delta=10{\rm K}$. 
The bottom panel  has the  same parameters as the middle one, 
except showing $-(T_c|_{\hat{\beta}(2T_m-T)}-T_c|_{\beta_m})/\Delta$
for the case of mirror transformed $\hat{\beta}$.
}
\label{fig_dns_bet_on_tc}
\end{figure}

\subsubsection{Test on the linearity of the $T_c -T_m$ shift}
\label{linearity test}
Looking at Tab. \ref{tab_dns_tc_tm} in more detail
we find that all changes in $(T_c-T_m)/\Delta$, which stem from 
the nonlinear $T$-dependence of the buoyancy force, i.e., from the differences between   
$\beta_m$ and $\hat{\beta}(T)$ (or $\hat{\beta}(2T_m-T)$) but having
the same temperature dependences of $\kappa$ and $\nu$, are comparable. 
To emphasize this feature, we look at the differences 
$T_c|_{\hat{\beta}(T)}-T_c|_{\beta_m}$ (or $-(T_c|_{\hat{\beta}(2T_m-T)}-T_c|_{\beta_m})$)
and plot them versus $\Delta$ as well as versus $Ra$, see figure \ref{fig_dns_bet_on_tc}.
We find a good collapse of the data onto a single curve 
for various temperature dependences of $\kappa$ and $\nu$.
In particular, the comparison between the middle and bottom panels of Fig. \ref{fig_dns_bet_on_tc}
leads to the relation $T_c|_{\hat{\beta}(T)}-T_c|_{\beta_m} =  -(T_c|_{\hat{\beta}(2T_m-T)}-T_c|_{\beta_m})$,
which indicates that the $T_c$ change is dominated by the quadratic term in $1-\rho(T)/\rho_m$,
but is almost independent of $\kappa$ and $\nu$.
This observation may be important for further attempts to improve the extended NOB BL theory.

\section{Conclusions}
\label{section/conclusions}

In this paper we have first presented in full detail the extension of boundary-layer theory to the case 
of compressible NOB fluids in a Rayleigh-B\'enard system. The theory predicts a deviation of the 
center temperature $T_c$ from the arithmetic mean temperature $T_m$ between the top and bottom plates, 
i.e., $T_c - T_m \neq 0$. 

Second, the theory has been tested against new experimental data for ethane near the critical point in its more
liquid-like phase. Data come from a series of experiments in cylindrical cells  of  aspect ratio $\Gamma = 0.5$ 
and $1$, reaching $Ra$ numbers $O(10^{10})$. The experimental measurements at $T_m = 27^{\circ}$C,  
$P/P_{*} = 1.062$ (and $\Gamma = 1$) have been chosen for comparison with those at $T_m = 40^{\circ}$C.
Contrary to the good agreement observed for the case of gas-like ethane \cite{ahl06}, the BL theory here gives 
much smaller values of the center temperature shift as experiment.

Third, direct numerical simulations DNS, based on $T$-dependent material parameters but still within 
the incompressible approximation and a two-dimensional domain have been performed to get more 
insight into the observed discrepancy between experiment 
and extended BL theory. The DNS results provide a satisfactory agreement with experiment both in 
the gas-like as well as in the liquid-like cases. 
Several hypothetical ethane-like fluids have been investigated too. Our analysis shows that the extended 
BL predictions fail whenever the non-linear temperature dependence of the density $\rho(T)$
is implemented in the numerical simulations. Furthermore, 
if the dependence of $\hat{\beta}(T)$ on $T$ dominates the NOB effects,
the sign of the linear term in the effective expansion function 
$\hat{\beta}(T)$ is  responsible for the sign of the variations of $T_c$ as compared to the OB value $T_c=T_m$.
But even if the difference $\hat{\beta}(T) - \beta_m$ is a non linear function of the temperature, the tiny $T_c$ 
shift detected in our simulations is proportional to $T_c|_{\hat{\beta}(T)}-T_c|_{\beta_m}$.

Our results can be summarized in the following physical picture of the surprising phenomenon that the center 
temperature $T_c$ in liquid-like ethane near its critical point increases, the center becomes warmer than 
the arithmeric mean temperature $T_m$, while in gas-like ethane the center gets colder, $T_c$ is less than 
$T_m$. Namely, in the liquid-like case the buoyancy $\propto (T-T_m) \hat{\beta}(T)$ is larger at the bottom
and smaller at the top, supporting the uprising warmer plumes more than the down coming colder top plumes.
This brings predominantly hotter material into the bulk. For gas-like ethane the buoyancy is larger at the 
cooler top, which accelerates the downgoing cold plumes with more preference than the uprising warm plumes 
from the bottom, that experience a weaker buoyancy. This in turn brings more cooler material into the bulk,
leading to $T_c < T_m$. It is the sign of the slope of $\hat{\beta}$ (or of $\beta$), which is the relevant
quantity.   

The more general lesson which can be drawn from this paper is that there
is a plethora of origins of NOB corrections. Which one dominates 
can only be said by having a detailed look at the temperature dependence
of all material parameters. Both the extended BL theory and 2D DNS are
useful tools to judge which temperature dependence is the most relevant
one or whether they all matter, as we now have often seen.

\section{acknowledgments}
This work is part of the research programme of the Stichting voor Fundamenteel 
Onderzoek der Materie (FOM), which is financially supported by the Nederlandse 
Organisatie voor Wetenschappelijk Onderzoek (NWO). 
The experimental work was supported by Grant DMR07-02111 of the US National 
Science Foundation.

\appendix

\section{Boundary-layer equations}
\label{section/appendix}

\subsection{Viscous boundary-layer}
\label{appendix/viscous-bl/2d}

Consider two-dimensional flow over a flat plate, such that the main 
wind velocity $U$ does not depend on $x$ and  $\lim_{z\to\infty}\ux(x,z) = U$. 
Then, the $x$-momentum near the wall is governed by \cite{sch00}
\begin{eqnarray}
\label{x-momentum/gas}
\rho
\left\{
\ux\,\duxdx + \uz\,\duxdz 
\right\}
&=&
\eta \, \left\{ \duxdxx + \duxdzz \right\}\\
&&
\nonumber
+ \detadz \, \left\{ \duxdz + \duzdx \right\}\\
&&
\nonumber
+
\ddx \left\{ \left[ \frac{\eta}{3} + \zeta \right] \left[ \duxdx + \duzdz \right] \right\},
\end{eqnarray}
where $\eta$ is the dynamic shear viscosity and $\zeta$ the volume expansion viscosity.

To appraise the dominant structure of equation (\ref{x-momentum/gas}) 
at sufficiently large Reynolds numbers, we follow Prandtl's scaling:
\begin{eqnarray}
\label{Pr-BL-scaling/x}
x 
&\;\;=\;\;& 
L\; \xtilde,\\
z 
&\;\;=\;\;& 
\frac{L}{\sqrt{\mbox{Re}}} \; \ztilde,\\
\ux 
&\;\;=\;\;& 
U \; \uxtilde,\\
\label{Pr-BL-scaling/uz}
\uz 
&\;\;=\;\;& 
\frac{U}{\sqrt{\mbox{Re}}} \; \uztilde,
\end{eqnarray}
where $L$ denotes the typical length of the plate in flow direction and $\mbox{Re} = L\,U\,\rho_m/\eta_m$ 
the Reynolds number. The index $m$ indicates that the fluid properties 
are evaluated at a thermodynamic reference state $(T_m, P_m, \rho_m)$, which is adopted  
in the nondimensionalization of $\rho \equiv \rho_m\,\rhotilde$ and $\eta \equiv \eta_{m}\etatilde$. 
Then, substituting (\ref{Pr-BL-scaling/x})-(\ref{Pr-BL-scaling/uz}) 
into (\ref{x-momentum/gas}), one finds
\begin{eqnarray}
\label{Prandtl/gas/Re}
\rhotilde
\left\{
\uxtilde\,\duxdxtilde + \uztilde\,\duxdztilde 
\right\}
&=&
\etatilde \, \left\{ \frac{1}{\mbox{Re}}\duxdxxtilde + \duxdzztilde \right\}\\
\nonumber
&&
+ 
\detadztilde \, \left\{ \duxdztilde + \frac{1}{\mbox{Re}}\duzdxtilde \right\}\\
\nonumber
&&
+
\frac{1}{\mbox{Re}}\ddxtilde \left\{ \etatilde_{0} \,
\left[ \duxdxtilde + \duzdztilde \right] \right\},
\end{eqnarray}
where $ \etatilde_{0} = \etatilde\left[ \frac{1}{3} + \frac{\zeta}{\eta} \right]$.
Clearly, all terms on the left-hand side of equation (\ref{Prandtl/gas/Re}) are 
of order unity. However, this is not the case on the right-hand side of 
(\ref{Prandtl/gas/Re}): only the viscous contributions involving 
\textit{transverse} gradients of $\ux$ are of order 1; the remaining 
terms are of order $1/\mbox{Re}$.\symbolfootnote[3]{Note that the term involving 
$\tilde{\nabla} \cdot \tilde{\mathbf{u}} = \duxdxtilde + \duzdztilde$
is of order $1/\mbox{Re}$, as long as $\zeta$ and $\eta$ are of the same order 
of magnitude. Indeed, $\zeta/\eta$ is of order unity if acoustical effects 
($\mbox{Ma} \ll 1$) and chemical reactions do not take place in the fluid.
For situations in which $\zeta > \eta$, see \cite[Section 81]{ll87}.}

Therefore, at large $\mbox{Re}$, the dominant part of the $x$-momentum equation 
is given by
\begin{eqnarray*}
\rhotilde
\left\{
\uxtilde\,\duxdxtilde + \uztilde\,\duxdztilde 
\right\}
&\;\;=\;\;&
\etatilde \, \duxdzztilde
+ \detadztilde \, \duxdztilde . 
\end{eqnarray*}

\subsection{Thermal boundary-layer}
\label{appendix/energy/gas}
Consider again two-dimensional, subsonic, and steady flow over a 
flat plate. In the framework of boundary-layer theory, energy 
conservation leads to the following equation for the entropy 
per mass $s$ \cite{ll87}:
\begin{equation}
\label{entropy/bl}
\ux \; \rho\, T \; \dsdx
\;+\;
\uz \, \rho\,T \; \dsdz 
\;\;=\;\;
\ddz \,\left\{ \Lambda\, \dTdz \right\}.
\end{equation}
Letting $s = s(T, \rho)$, the entropy gradient (say, its $\dz s$ component) 
can be expressed as
\[
\dsdz
\;\;=\;\;
\dsdT \, \dTdz \;+\; \dsdrho \, \drhodz.
\]
The first contribution is directly associated with the 
isochoric specific heat (per mass) of the gas
\[
T\,\dsdT 
\;\equiv\;
c_V
\;\equiv\;
\frac{c_P}{\gamma}.
\]
The second contribution follows from a Maxwell relation,
\[
\dsdrho 
\;\;=\;\; 
-\frac{1}{\rho^2} \, \dPdT
\;\;=\;\;
-\frac{1}{\rho\,T}\,\frac{c_P}{\gamma}\,\frac{\gamma -1}{\beta}.
\]   
Thus, the left-hand side of equation (\ref{entropy/bl}) can 
be written as
\begin{eqnarray*}
\ux \; \rho\, T \; \dsdx
\;+\;
\uz \, \rho\,T \; \dsdz 
&=&
\frac{\rho\,c_P}{\gamma}
\left\{\ux\,\dTdx + \uz \, \dTdz \right\}\\
&&
-
\frac{c_P}{\gamma} \frac{\gamma-1}{\beta}\,
\left\{\ux\,\drhodx + \uz \, \drhodz 
\right\} .
\end{eqnarray*}
Finally, using the continuity equation 
$\ux \, \dx\rho + \uz \, \dz\rho = -\rho \, \{ \dx \ux + \dz \uz \}$,  
one finds
\[
\ux\,\dTdx + \uz \, \dTdz   
+
\frac{\gamma - 1}{\beta}
\left\{ \duxdx + \duzdz \right\}
=
\frac{\gamma}{\rho\,c_P}
\ddz\left\{\Lambda\,\dTdz\right\}
.
\]
Note that the limiting case of liquids (namely $\gamma = 1$) 
is fully accounted by this equation.

\section{Self-similarity Ansatz}
\label{appb}

\subsection{Viscous boundary-layer}
\label{appendix/viscous-bl/gas}

In the stream-function representation (\ref{rhotilde.ux}), 
the longitudinal velocity $\ux$ is expressed as
\[
\ux 
\;\;=\;\;
\frac{1}{\rhotilde}
\ddz \{\lc\,\Uc\,\Psitilde \}
\;\;=\;\; 
\Uc\, \frac{\Psitilde\,'}{\rhotilde}
\]
and its spatial derivatives are
\begin{eqnarray}
\duxdx
\label{dx_ux/gas}
&\;\;=\;\;&
-\frac{\nu_m}{2\lc^2}\; 
\frac{\ztildex}{\rhotilde^2} 
\left\{
\rhotilde\,\Psitilde\,'' 
- 
\rhotilde\,'\,\Psitilde\,' 
\right \},\\
\duxdz
\label{dz_ux/gas}
&\;\;=\;\;&
\frac{\Uc}{\lc}\; 
\frac{1}{\rhotilde^2} 
\left\{
\rhotilde\,\Psitilde\,'' 
- 
\rhotilde\,'\,\Psitilde\,' 
\right \}
,\\
\nonumber
\duxdzz 
&\;\;=\;\;&
\frac{\Uc}{\lc^2}
\frac{1}{\rhotilde^3}
\left\{
\rhotilde^2\,\Psitilde\,''' 
-
2\rhotilde\,\rhotilde\,'\,\Psitilde\,''
\right.\\
&&
\label{dzz_ux/gas}
\qquad \;\;\;\;\;\;
+
\left.
\left[
2(\rhotilde\,')^2 - \rhotilde\,\rhotilde\,''
\right]
\Psitilde\,'
\right\}.
\end{eqnarray}

Likewise, from equation (\ref{rhotilde.uz}), the transverse velocity $\uz$ reads
\[
\uz 
\;\;=\;\; 
-\, \frac{1}{\rhotilde}\ddx \{ \Uc \,\lc \,\Psitilde \}
\;\;=\;\;
\frac{\nu_{m}}{2\lc}
\left\{ 
\ztildex\,\frac{\Psitilde\,'}{\rhotilde} - \frac{\Psitilde}{\rhotilde}
\right\},
\]
with
\begin{eqnarray}
\label{dz_uz/gas}
\duzdz 
\;\;=\;\;
\frac{\nu_{m}}{2\lc^2}
\frac{1}{\rhotilde^2}
\left\{ 
\ztildex\,\rhotilde\,\Psitilde\,''
- 
\ztildex\,\rhotilde\,'\,\Psitilde\,'
+
\rhotilde\,'\,\Psitilde
\right\}.
\end{eqnarray}
Thus, the advective contributions in Prandtl's equation (\ref{Prandtl}) can 
be written as
\begin{eqnarray}
\rho\left\{\ux \duxdx + \uz \duxdz \right\}
\nonumber
&\;\;=\;\;&
-\frac{1}{2}
\frac{\eta_m\,\Uc}{\rhotilde\,\lc^2}\,
\left\{
\Psitilde\,\Psitilde\,'' 
\right. \qquad\\
\label{advection/gas}
&&
\qquad \;\;
-
\left.
\frac{\rhotilde\,'}{\rhotilde}\,\Psitilde\,\Psitilde\,' 
\right \}.
\end{eqnarray}

Now consider the viscous contributions:
\begin{equation}
\label{viscous-contributions}
\ddz \left\{ \eta \duxdz \right\}
\;\;=\;\;
\eta\,\duxdzz  \;+\; \detadz \; \duxdz.
\end{equation}
Since
\begin{equation}
\label{d-eta-tilde}
\detadz 
\;\;=\;\; 
\frac{\eta_{m}}{\lc}\,
\left\{
\detadtheta\,\Thetatilde\,'
+
\detadrho\,\rhotilde\,'
\right\},
\end{equation}
insertion of (\ref{dz_ux/gas}) and (\ref{dzz_ux/gas}) into 
(\ref{viscous-contributions}) leads to
\begin{eqnarray}
\nonumber
\frac{\rhotilde\,\lc^2}{\eta_m\,\Uc}
\ddz \left\{ \eta \duxdz \right\}
&=&
\etatilde\,\Psitilde\,''' 
+
\left[
\etatilde\,'
-
2\frac{\rhotilde\,'}{\rhotilde}\,\etatilde
\right]\,\Psitilde\,''\\
\nonumber
&&
+
\left\{
\left[
2\left(\frac{\rhotilde\,'}{\rhotilde}\right)^{2} 
- \frac{\rhotilde\,''}{\rhotilde}
\right]\etatilde \right.\\
&&
\label{viscous/gas}
\qquad \qquad
-
\left.
\frac{\rhotilde\,'}{\rhotilde}\,\etatilde\,'
\right\} \Psitilde\,'.
\end{eqnarray}

Therefore, by substituting (\ref{advection/gas}) and (\ref{viscous/gas}) 
into (\ref{Prandtl}) one finally obtains equation (\ref{Blasius}).

\subsection{Thermal boundary-layer}
\label{appendix/thermal-bl/gas}

In the same spirit as above, inserting (\ref{dx_ux/gas}), (\ref{dz_uz/gas}), and 
\[
\ddz\left\{ \Lambda \,\dTdz\right\} 
\;\;=\;\; 
\frac{\Lambda_m\,\Delta}{\lc^2}
\left\{ \Lambdatilde\,\Thetatilde\,'' + \Lambdatilde\,'\,\Thetatilde\,' \right\}
\]
into equation (\ref{temperature/NOB/gas}) one finds
\begin{eqnarray*}
-\frac{\nu_m\,\Delta}{2\lc^2} \frac{\Psitilde}{\rhotilde}\Thetatilde\,'
+
\frac{\nu_m}{2\lc^2}
\frac{\gamma-1}{\beta}
\frac{\rhotilde\,'}{\rhotilde^2}\,\Psitilde
&=&
\frac{\gamma}{\rhotilde\,\cptilde} \,
\frac{1}{\Pr}
\left\{ \Lambdatilde\,\Thetatilde\,'' + \Lambdatilde\,'\,\Thetatilde\,' \right\},
\end{eqnarray*}
where $\betatilde = \beta\,\Delta$ and $\Pr = (\nu_m\,\rho_m\,c_{P,m})/\Lambda_{m}$.
Thus
\[
\Lambdatilde\,\Thetatilde\,'' 
\;+\;
\left\{
\frac{1}{2} \frac{\cptilde}{\gamma} \, \Pr\, \Psitilde
+
\Lambdatilde\,'
\right\}\,\Thetatilde\,'
\;- \;
\frac{\gamma-1}{2\gamma}\,
\frac{\cptilde}{\betatilde}\,
\frac{\rhotilde\,'}{\rhotilde}\,\Pr\,\Psitilde
=
0.
\]
Here, substituting $\rhotilde\,'$ by (\ref{rhotilde/slope}) one 
finally obtains
\[
\Lambdatilde\,\Thetatilde\,'' 
\;+\;
\left\{
\frac{1}{2}\,\cptilde \, \Pr\, \Psitilde
\;+\;
\Lambdatilde\,'
\right\}\,\Thetatilde\,'
\;\;=\;\;
0.
\]
Note that the limiting case of liquids (namely $\gamma = 1$) 
is fully accounted by this equation.

\section{Numerical results on Nusselt numbers in real and hypothetical ethane fluids}
\label{appc}
For completeness, in table \ref{tab_dns_nu_dev} the NOB corrections in the Nusselt number are
given, resulting from the numerical simulations of real and hypothetical ethane. The corresponding
NOB corrections of the center temperature had already been shown in table \ref{tab_dns_tc_tm}.

\begin{table*}

\begin{center}
{\small
\begin{tabular}{llllrrr}
\hline
case&$\hat{\beta}$&$\kappa$&$\nu$&$100(Nu/Nu_{OB}-1)$&$100(Nu/Nu_{OB}-1)$&$100(Nu/Nu_{OB}-1)$\\
&&&&at $Ra=10^6$&at $Ra=10^7$&at $Ra=10^8$\\\hline
1 (NOB)&$\hat{\beta}(T)$     &$\kappa(T)$&$\nu(T)$&$ 0.7097\pm 0.1517$&$ 0.8956\pm 0.2213$&$ 1.2403\pm 0.2316$\\
2      &$\hat{\beta}(T)$     &$\kappa(T)$&$\nu_m$ &$ 0.1655\pm 0.1511$&$ 0.4795\pm 0.2323$&$ 0.4040\pm 0.2328$\\
3      &$\hat{\beta}(T)$     &$\kappa_m$ &$\nu(T)$&$ 1.3160\pm 0.1518$&$ 1.5812\pm 0.2230$&$ 1.6502\pm 0.2234$\\
4      &$\hat{\beta}(T)$     &$\kappa_m$ &$\nu_m$ &$ 0.6640\pm 0.1505$&$ 0.4963\pm 0.2324$&$ 0.6450\pm 0.2425$\\
5      &$\hat{\beta}(2T_m-T)$&$\kappa(T)$&$\nu(T)$&$ 1.0087\pm 0.1518$&$ 1.2804\pm 0.2214$&$ 1.3223\pm 0.2228$\\
6      &$\hat{\beta}(2T_m-T)$&$\kappa(T)$&$\nu_m$ &$ 0.1655\pm 0.1511$&$ 0.4795\pm 0.2323$&$ 0.4040\pm 0.2328$\\
7      &$\hat{\beta}(2T_m-T)$&$\kappa_m$ &$\nu(T)$&$ 0.9925\pm 0.1518$&$ 1.1741\pm 0.2255$&$ 1.0832\pm 0.2329$\\
8      &$\hat{\beta}(2T_m-T)$&$\kappa_m$ &$\nu_m$ &$ 0.5988\pm 0.1505$&$ 0.9991\pm 0.2236$&$ 1.0073\pm 0.2286$\\
9      &$\beta_m$            &$\kappa(T)$&$\nu(T)$&$ 0.3425\pm 0.1517$&$ 0.6188\pm 0.2175$&$ 0.0538\pm 0.2455$\\
10     &$\beta_m$            &$\kappa(T)$&$\nu_m$ &$-0.1734\pm 0.1509$&$-0.4404\pm 0.2277$&$ 0.0195\pm 0.2246$\\
11     &$\beta_m$            &$\kappa_m$ &$\nu(T)$&$ 0.6380\pm 0.1523$&$ 0.4491\pm 0.2321$&$ 0.7775\pm 0.2222$\\
12 (OB)&$\beta_m$            &$\kappa_m$ &$\nu_m$ &$ 0.0000\pm 0.1070$&$ 0.0000\pm 0.1573$&$ 0.0000\pm 0.1689$\\
\hline
\end{tabular}
}
\end{center}

\caption{ 
Simulation results of the relative deviation of the Nusselt number $Nu/Nu_{OB}-1$ 
at the temperature difference of $\Delta=10$K for several hypothetical fluids. 
At the Rayleigh numbers of $Ra=10^6$, $10^7$, and $10^8$ 
the OB Nusselt numbers correspond to $Nu_{OB}=6.53$, $12.43$, and $25.12$, respectively.
The notations for the material properties are the same as in Table \ref{tab_dns_tc_tm}.
}
\label{tab_dns_nu_dev}
\end{table*}




\begin{thebibliography}{60}
\expandafter\ifx\csname natexlab\endcsname\relax\def\natexlab#1{#1}\fi
\expandafter\ifx\csname bibnamefont\endcsname\relax
  \def\bibnamefont#1{#1}\fi
\expandafter\ifx\csname bibfnamefont\endcsname\relax
  \def\bibfnamefont#1{#1}\fi
\expandafter\ifx\csname citenamefont\endcsname\relax
  \def\citenamefont#1{#1}\fi
\expandafter\ifx\csname url\endcsname\relax
  \def\url#1{\texttt{#1}}\fi
\expandafter\ifx\csname urlprefix\endcsname\relax\def\urlprefix{URL }\fi
\providecommand{\bibinfo}[2]{#2}
\providecommand{\eprint}[2][]{\url{#2}}

\bibitem[{\citenamefont{Castaing et~al.}(1989)\citenamefont{Castaing,
  Gunaratne, Heslot, Kadanoff, Libchaber, Thomae, Wu, Zaleski, and
  Zanetti}}]{cas89}
\bibinfo{author}{\bibfnamefont{B.}~\bibnamefont{Castaing}},
  \bibinfo{author}{\bibfnamefont{G.}~\bibnamefont{Gunaratne}},
  \bibinfo{author}{\bibfnamefont{F.}~\bibnamefont{Heslot}},
  \bibinfo{author}{\bibfnamefont{L.}~\bibnamefont{Kadanoff}},
  \bibinfo{author}{\bibfnamefont{A.}~\bibnamefont{Libchaber}},
  \bibinfo{author}{\bibfnamefont{S.}~\bibnamefont{Thomae}},
  \bibinfo{author}{\bibfnamefont{X.~Z.} \bibnamefont{Wu}},
  \bibinfo{author}{\bibfnamefont{S.}~\bibnamefont{Zaleski}}, \bibnamefont{and}
  \bibinfo{author}{\bibfnamefont{G.}~\bibnamefont{Zanetti}},
  \bibinfo{journal}{J. Fluid Mech.} \textbf{\bibinfo{volume}{204}},
  \bibinfo{pages}{1} (\bibinfo{year}{1989}).

\bibitem[{\citenamefont{Siggia}(1994)}]{sig94}
\bibinfo{author}{\bibfnamefont{E.~D.} \bibnamefont{Siggia}},
  \bibinfo{journal}{Annu. Rev. Fluid Mech.} \textbf{\bibinfo{volume}{26}},
  \bibinfo{pages}{137} (\bibinfo{year}{1994}).

\bibitem[{\citenamefont{Cioni et~al.}(1997)\citenamefont{Cioni, Ciliberto, and
  Sommeria}}]{cio97}
\bibinfo{author}{\bibfnamefont{S.}~\bibnamefont{Cioni}},
  \bibinfo{author}{\bibfnamefont{S.}~\bibnamefont{Ciliberto}},
  \bibnamefont{and} \bibinfo{author}{\bibfnamefont{J.}~\bibnamefont{Sommeria}},
  \bibinfo{journal}{J. Fluid Mech.} \textbf{\bibinfo{volume}{335}},
  \bibinfo{pages}{111} (\bibinfo{year}{1997}).

\bibitem[{\citenamefont{Chavanne et~al.}(1997)\citenamefont{Chavanne, Chilla,
  Castaing, Hebral, Chabaud, and Chaussy}}]{cha97}
\bibinfo{author}{\bibfnamefont{X.}~\bibnamefont{Chavanne}},
  \bibinfo{author}{\bibfnamefont{F.}~\bibnamefont{Chilla}},
  \bibinfo{author}{\bibfnamefont{B.}~\bibnamefont{Castaing}},
  \bibinfo{author}{\bibfnamefont{B.}~\bibnamefont{Hebral}},
  \bibinfo{author}{\bibfnamefont{B.}~\bibnamefont{Chabaud}}, \bibnamefont{and}
  \bibinfo{author}{\bibfnamefont{J.}~\bibnamefont{Chaussy}},
  \bibinfo{journal}{Phys. Rev. Lett.} \textbf{\bibinfo{volume}{79}},
  \bibinfo{pages}{3648} (\bibinfo{year}{1997}).

\bibitem[{\citenamefont{Xu et~al.}(2000)\citenamefont{Xu, Bajaj, and
  Ahlers}}]{xu00}
\bibinfo{author}{\bibfnamefont{X.}~\bibnamefont{Xu}},
  \bibinfo{author}{\bibfnamefont{K.~M.~S.} \bibnamefont{Bajaj}},
  \bibnamefont{and} \bibinfo{author}{\bibfnamefont{G.}~\bibnamefont{Ahlers}},
  \bibinfo{journal}{Phys. Rev. Lett.} \textbf{\bibinfo{volume}{84}},
  \bibinfo{pages}{4357} (\bibinfo{year}{2000}).

\bibitem[{\citenamefont{Niemela et~al.}(2000)\citenamefont{Niemela, Skrebek,
  Sreenivasan, and Donnelly}}]{nie00}
\bibinfo{author}{\bibfnamefont{J.}~\bibnamefont{Niemela}},
  \bibinfo{author}{\bibfnamefont{L.}~\bibnamefont{Skrebek}},
  \bibinfo{author}{\bibfnamefont{K.~R.} \bibnamefont{Sreenivasan}},
  \bibnamefont{and} \bibinfo{author}{\bibfnamefont{R.}~\bibnamefont{Donnelly}},
  \bibinfo{journal}{Nature} \textbf{\bibinfo{volume}{404}},
  \bibinfo{pages}{837} (\bibinfo{year}{2000}).

\bibitem[{\citenamefont{Chavanne et~al.}(2001)\citenamefont{Chavanne, Chilla,
  Chabaud, Castaing, and Hebral}}]{cha01}
\bibinfo{author}{\bibfnamefont{X.}~\bibnamefont{Chavanne}},
  \bibinfo{author}{\bibfnamefont{F.}~\bibnamefont{Chilla}},
  \bibinfo{author}{\bibfnamefont{B.}~\bibnamefont{Chabaud}},
  \bibinfo{author}{\bibfnamefont{B.}~\bibnamefont{Castaing}}, \bibnamefont{and}
  \bibinfo{author}{\bibfnamefont{B.}~\bibnamefont{Hebral}},
  \bibinfo{journal}{Phys. Fluids} \textbf{\bibinfo{volume}{13}},
  \bibinfo{pages}{1300} (\bibinfo{year}{2001}).

\bibitem[{\citenamefont{Ahlers and Xu}(2001)}]{ahl01}
\bibinfo{author}{\bibfnamefont{G.}~\bibnamefont{Ahlers}} \bibnamefont{and}
  \bibinfo{author}{\bibfnamefont{X.}~\bibnamefont{Xu}}, \bibinfo{journal}{Phys.
  Rev. Lett.} \textbf{\bibinfo{volume}{86}}, \bibinfo{pages}{3320}
  (\bibinfo{year}{2001}).

\bibitem[{\citenamefont{Qiu and Tong}(2001)}]{qiu01b}
\bibinfo{author}{\bibfnamefont{X.~L.} \bibnamefont{Qiu}} \bibnamefont{and}
  \bibinfo{author}{\bibfnamefont{P.}~\bibnamefont{Tong}},
  \bibinfo{journal}{Phys. Rev. E} \textbf{\bibinfo{volume}{64}},
  \bibinfo{pages}{036304} (\bibinfo{year}{2001}).

\bibitem[{\citenamefont{Kadanoff}(2001)}]{kad01}
\bibinfo{author}{\bibfnamefont{L.~P.} \bibnamefont{Kadanoff}},
  \bibinfo{journal}{Phys. Today} \textbf{\bibinfo{volume}{54}},
  \bibinfo{pages}{34} (\bibinfo{year}{2001}).

\bibitem[{\citenamefont{Xia et~al.}(2002)\citenamefont{Xia, Lam, and
  Zhou}}]{xia02}
\bibinfo{author}{\bibfnamefont{K.-Q.} \bibnamefont{Xia}},
  \bibinfo{author}{\bibfnamefont{S.}~\bibnamefont{Lam}}, \bibnamefont{and}
  \bibinfo{author}{\bibfnamefont{S.~Q.} \bibnamefont{Zhou}},
  \bibinfo{journal}{Phys. Rev. Lett.} \textbf{\bibinfo{volume}{88}},
  \bibinfo{pages}{064501} (\bibinfo{year}{2002}).

\bibitem[{\citenamefont{Roche et~al.}(2002)\citenamefont{Roche, Castaing,
  Chabaud, and Hebral}}]{roc02}
\bibinfo{author}{\bibfnamefont{P.~E.} \bibnamefont{Roche}},
  \bibinfo{author}{\bibfnamefont{B.}~\bibnamefont{Castaing}},
  \bibinfo{author}{\bibfnamefont{B.}~\bibnamefont{Chabaud}}, \bibnamefont{and}
  \bibinfo{author}{\bibfnamefont{B.}~\bibnamefont{Hebral}},
  \bibinfo{journal}{Europhys. Lett.} \textbf{\bibinfo{volume}{58}},
  \bibinfo{pages}{693} (\bibinfo{year}{2002}).

\bibitem[{\citenamefont{Niemela and Sreenivasan}(2003)}]{nie03}
\bibinfo{author}{\bibfnamefont{J.}~\bibnamefont{Niemela}} \bibnamefont{and}
  \bibinfo{author}{\bibfnamefont{K.~R.} \bibnamefont{Sreenivasan}},
  \bibinfo{journal}{J. Fluid Mech.} \textbf{\bibinfo{volume}{481}},
  \bibinfo{pages}{355} (\bibinfo{year}{2003}).

\bibitem[{\citenamefont{Funfschilling and Ahlers}(2004)}]{fun04}
\bibinfo{author}{\bibfnamefont{D.}~\bibnamefont{Funfschilling}}
  \bibnamefont{and} \bibinfo{author}{\bibfnamefont{G.}~\bibnamefont{Ahlers}},
  \bibinfo{journal}{Phys. Rev. Lett.} \textbf{\bibinfo{volume}{92}},
  \bibinfo{pages}{194502} (\bibinfo{year}{2004}).

\bibitem[{\citenamefont{Brown et~al.}(2005{\natexlab{a}})\citenamefont{Brown,
  Funfschilling, and Ahlers}}]{bro05b}
\bibinfo{author}{\bibfnamefont{E.}~\bibnamefont{Brown}},
  \bibinfo{author}{\bibfnamefont{D.}~\bibnamefont{Funfschilling}},
  \bibnamefont{and} \bibinfo{author}{\bibfnamefont{G.}~\bibnamefont{Ahlers}},
  \bibinfo{journal}{Phys. Rev. Lett.} \textbf{\bibinfo{volume}{95}},
  \bibinfo{pages}{084503} (\bibinfo{year}{2005}{\natexlab{a}}).

\bibitem[{\citenamefont{Nikolaenko et~al.}(2005)\citenamefont{Nikolaenko,
  Brown, Funfschilling, and Ahlers}}]{nik05}
\bibinfo{author}{\bibfnamefont{A.}~\bibnamefont{Nikolaenko}},
  \bibinfo{author}{\bibfnamefont{E.}~\bibnamefont{Brown}},
  \bibinfo{author}{\bibfnamefont{D.}~\bibnamefont{Funfschilling}},
  \bibnamefont{and} \bibinfo{author}{\bibfnamefont{G.}~\bibnamefont{Ahlers}},
  \bibinfo{journal}{J. Fluid Mech.} \textbf{\bibinfo{volume}{523}},
  \bibinfo{pages}{251} (\bibinfo{year}{2005}).

\bibitem[{\citenamefont{Niemela and Sreenivasan}(2006{\natexlab{a}})}]{nie06}
\bibinfo{author}{\bibfnamefont{J.}~\bibnamefont{Niemela}} \bibnamefont{and}
  \bibinfo{author}{\bibfnamefont{K.~R.} \bibnamefont{Sreenivasan}},
  \bibinfo{journal}{J. Fluid Mech.} \textbf{\bibinfo{volume}{557}},
  \bibinfo{pages}{411 } (\bibinfo{year}{2006}{\natexlab{a}}).

\bibitem[{\citenamefont{Xia et~al.}(2003)\citenamefont{Xia, Sun, and
  Zhou}}]{xia03}
\bibinfo{author}{\bibfnamefont{K.-Q.} \bibnamefont{Xia}},
  \bibinfo{author}{\bibfnamefont{C.}~\bibnamefont{Sun}}, \bibnamefont{and}
  \bibinfo{author}{\bibfnamefont{S.~Q.} \bibnamefont{Zhou}},
  \bibinfo{journal}{Phys. Rev. E} \textbf{\bibinfo{volume}{68}},
  \bibinfo{pages}{066303} (\bibinfo{year}{2003}).

\bibitem[{\citenamefont{Shang et~al.}(2003)\citenamefont{Shang, Qiu, Tong, and
  Xia}}]{sha03}
\bibinfo{author}{\bibfnamefont{X.~D.} \bibnamefont{Shang}},
  \bibinfo{author}{\bibfnamefont{X.~L.} \bibnamefont{Qiu}},
  \bibinfo{author}{\bibfnamefont{P.}~\bibnamefont{Tong}}, \bibnamefont{and}
  \bibinfo{author}{\bibfnamefont{K.-Q.} \bibnamefont{Xia}},
  \bibinfo{journal}{Phys. Rev. Lett.} \textbf{\bibinfo{volume}{90}},
  \bibinfo{pages}{074501} (\bibinfo{year}{2003}).

\bibitem[{\citenamefont{Roche et~al.}(2004)\citenamefont{Roche, Castaing,
  Chabaud, and Hebral}}]{roc04}
\bibinfo{author}{\bibfnamefont{P.~E.} \bibnamefont{Roche}},
  \bibinfo{author}{\bibfnamefont{B.}~\bibnamefont{Castaing}},
  \bibinfo{author}{\bibfnamefont{B.}~\bibnamefont{Chabaud}}, \bibnamefont{and}
  \bibinfo{author}{\bibfnamefont{B.}~\bibnamefont{Hebral}},
  \bibinfo{journal}{J. Low. Temp. Phys.} \textbf{\bibinfo{volume}{134}},
  \bibinfo{pages}{1011} (\bibinfo{year}{2004}).

\bibitem[{\citenamefont{Sun et~al.}(2005)\citenamefont{Sun, Xia, and
  Tong}}]{sun05}
\bibinfo{author}{\bibfnamefont{C.}~\bibnamefont{Sun}},
  \bibinfo{author}{\bibfnamefont{K.~Q.} \bibnamefont{Xia}}, \bibnamefont{and}
  \bibinfo{author}{\bibfnamefont{P.}~\bibnamefont{Tong}},
  \bibinfo{journal}{Phys. Rev. E} \textbf{\bibinfo{volume}{72}},
  \bibinfo{pages}{026302} (\bibinfo{year}{2005}).

\bibitem[{\citenamefont{Brown and Ahlers}(2006)}]{bro06}
\bibinfo{author}{\bibfnamefont{E.}~\bibnamefont{Brown}} \bibnamefont{and}
  \bibinfo{author}{\bibfnamefont{G.}~\bibnamefont{Ahlers}},
  \bibinfo{journal}{J. Fluid Mech.} \textbf{\bibinfo{volume}{568}},
  \bibinfo{pages}{351} (\bibinfo{year}{2006}).

\bibitem[{\citenamefont{Niemela and Sreenivasan}(2006{\natexlab{b}})}]{nie06b}
\bibinfo{author}{\bibfnamefont{J.}~\bibnamefont{Niemela}} \bibnamefont{and}
  \bibinfo{author}{\bibfnamefont{K.~R.} \bibnamefont{Sreenivasan}},
  \bibinfo{journal}{J. Low Temp. Phys.} \textbf{\bibinfo{volume}{143}},
  \bibinfo{pages}{163 } (\bibinfo{year}{2006}{\natexlab{b}}).

\bibitem[{\citenamefont{du~Puits et~al.}(2007)\citenamefont{du~Puits, Resagk,
  Tilgner, Busse, and Thess}}]{pui07}
\bibinfo{author}{\bibfnamefont{R.}~\bibnamefont{du~Puits}},
  \bibinfo{author}{\bibfnamefont{C.}~\bibnamefont{Resagk}},
  \bibinfo{author}{\bibfnamefont{A.}~\bibnamefont{Tilgner}},
  \bibinfo{author}{\bibfnamefont{F.~H.} \bibnamefont{Busse}}, \bibnamefont{and}
  \bibinfo{author}{\bibfnamefont{A.}~\bibnamefont{Thess}}, \bibinfo{journal}{J.
  Fluid Mech.} \textbf{\bibinfo{volume}{572}}, \bibinfo{pages}{231}
  (\bibinfo{year}{2007}).

\bibitem[{\citenamefont{Kerr}(1996)}]{ker96}
\bibinfo{author}{\bibfnamefont{R.}~\bibnamefont{Kerr}}, \bibinfo{journal}{J.
  Fluid Mech.} \textbf{\bibinfo{volume}{310}}, \bibinfo{pages}{139}
  (\bibinfo{year}{1996}).

\bibitem[{\citenamefont{Benzi et~al.}(1998)\citenamefont{Benzi, Toschi, and
  Tripiccione}}]{ben98}
\bibinfo{author}{\bibfnamefont{R.}~\bibnamefont{Benzi}},
  \bibinfo{author}{\bibfnamefont{F.}~\bibnamefont{Toschi}}, \bibnamefont{and}
  \bibinfo{author}{\bibfnamefont{R.}~\bibnamefont{Tripiccione}},
  \bibinfo{journal}{J. Stat. Phys.} \textbf{\bibinfo{volume}{93}},
  \bibinfo{pages}{901} (\bibinfo{year}{1998}).

\bibitem[{\citenamefont{Kerr and Herring}(2000)}]{ker00}
\bibinfo{author}{\bibfnamefont{R.}~\bibnamefont{Kerr}} \bibnamefont{and}
  \bibinfo{author}{\bibfnamefont{J.~R.} \bibnamefont{Herring}},
  \bibinfo{journal}{J. Fluid Mech.} \textbf{\bibinfo{volume}{419}},
  \bibinfo{pages}{325} (\bibinfo{year}{2000}).

\bibitem[{\citenamefont{Verzicco and Camussi}(1999)}]{ver99}
\bibinfo{author}{\bibfnamefont{R.}~\bibnamefont{Verzicco}} \bibnamefont{and}
  \bibinfo{author}{\bibfnamefont{R.}~\bibnamefont{Camussi}},
  \bibinfo{journal}{J. Fluid Mech.} \textbf{\bibinfo{volume}{383}},
  \bibinfo{pages}{55} (\bibinfo{year}{1999}).

\bibitem[{\citenamefont{Verzicco and Camussi}(2003)}]{ver03}
\bibinfo{author}{\bibfnamefont{R.}~\bibnamefont{Verzicco}} \bibnamefont{and}
  \bibinfo{author}{\bibfnamefont{R.}~\bibnamefont{Camussi}},
  \bibinfo{journal}{J. Fluid Mech.} \textbf{\bibinfo{volume}{477}},
  \bibinfo{pages}{19} (\bibinfo{year}{2003}).

\bibitem[{\citenamefont{Lohse and Toschi}(2003)}]{loh03}
\bibinfo{author}{\bibfnamefont{D.}~\bibnamefont{Lohse}} \bibnamefont{and}
  \bibinfo{author}{\bibfnamefont{F.}~\bibnamefont{Toschi}},
  \bibinfo{journal}{Phys. Rev. Lett.} \textbf{\bibinfo{volume}{90}},
  \bibinfo{pages}{034502} (\bibinfo{year}{2003}).

\bibitem[{\citenamefont{Verzicco}(2004)}]{ver04}
\bibinfo{author}{\bibfnamefont{R.}~\bibnamefont{Verzicco}},
  \bibinfo{journal}{Phys. Fluids} \textbf{\bibinfo{volume}{16}},
  \bibinfo{pages}{1965} (\bibinfo{year}{2004}).

\bibitem[{\citenamefont{Amati et~al.}(2005)\citenamefont{Amati, Koal,
  Massaioli, Sreenivasan, and Verzicco}}]{ama05}
\bibinfo{author}{\bibfnamefont{G.}~\bibnamefont{Amati}},
  \bibinfo{author}{\bibfnamefont{K.}~\bibnamefont{Koal}},
  \bibinfo{author}{\bibfnamefont{F.}~\bibnamefont{Massaioli}},
  \bibinfo{author}{\bibfnamefont{K.~R.} \bibnamefont{Sreenivasan}},
  \bibnamefont{and} \bibinfo{author}{\bibfnamefont{R.}~\bibnamefont{Verzicco}},
  \bibinfo{journal}{Phys. Fluids} \textbf{\bibinfo{volume}{17}},
  \bibinfo{pages}{121701} (\bibinfo{year}{2005}).

\bibitem[{\citenamefont{Shishkina and Wagner}(19)}]{shi06}
\bibinfo{author}{\bibfnamefont{O.}~\bibnamefont{Shishkina}} \bibnamefont{and}
  \bibinfo{author}{\bibfnamefont{C.}~\bibnamefont{Wagner}},
  \bibinfo{journal}{J. Fluid Mech.} \textbf{\bibinfo{volume}{546}},
  \bibinfo{pages}{51 } (\bibinfo{year}{19}).

\bibitem[{\citenamefont{Stringano and Verzicco}(2006)}]{str06}
\bibinfo{author}{\bibfnamefont{G.}~\bibnamefont{Stringano}} \bibnamefont{and}
  \bibinfo{author}{\bibfnamefont{R.}~\bibnamefont{Verzicco}},
  \bibinfo{journal}{J. Fluid Mech.} \textbf{\bibinfo{volume}{548}},
  \bibinfo{pages}{1} (\bibinfo{year}{2006}).

\bibitem[{\citenamefont{Kunnen et~al.}(2008)\citenamefont{Kunnen, Clercx,
  Geurts, Bokhoven, Akkermans, and Verzicco}}]{kun08}
\bibinfo{author}{\bibfnamefont{R.~P.~J.} \bibnamefont{Kunnen}},
  \bibinfo{author}{\bibfnamefont{H.~J.~H.} \bibnamefont{Clercx}},
  \bibinfo{author}{\bibfnamefont{B.~J.} \bibnamefont{Geurts}},
  \bibinfo{author}{\bibfnamefont{L.~A.} \bibnamefont{Bokhoven}},
  \bibinfo{author}{\bibfnamefont{R.~A.~D.} \bibnamefont{Akkermans}},
  \bibnamefont{and} \bibinfo{author}{\bibfnamefont{R.}~\bibnamefont{Verzicco}},
  \bibinfo{journal}{Phys. Rev. E} \textbf{77}, 016302 (\bibinfo{year}{2008}).

\bibitem[{\citenamefont{Grossmann and Lohse}(2004)}]{gro-all}
\bibinfo{author}{\bibfnamefont{S.}~\bibnamefont{Grossmann}} \bibnamefont{and}
  \bibinfo{author}{\bibfnamefont{D.}~\bibnamefont{Lohse}}, \bibinfo{journal}{J.
  Fluid Mech. \textbf{407}, 27 (2000). Phys. Rev. Lett. \textbf{86}, 3316
  (2001). Phys. Rev. E \textbf{66}, 016305 (2002). Phys. Fluids \textbf{16},
  4462}  (\bibinfo{year}{2004}).

\bibitem[{\citenamefont{Benzi}(2005)}]{ben05}
\bibinfo{author}{\bibfnamefont{R.}~\bibnamefont{Benzi}},
  \bibinfo{journal}{Phys. Rev. Lett.} \textbf{\bibinfo{volume}{95}},
  \bibinfo{pages}{024502} (\bibinfo{year}{2005}).

\bibitem[{\citenamefont{Brown and Ahlers}(2007{\natexlab{a}})}]{bro07}
\bibinfo{author}{\bibfnamefont{E.}~\bibnamefont{Brown}} \bibnamefont{and}
  \bibinfo{author}{\bibfnamefont{G.}~\bibnamefont{Ahlers}},
  \bibinfo{journal}{Phys. Rev. Lett.} \textbf{\bibinfo{volume}{98}},
  \bibinfo{pages}{134501} (\bibinfo{year}{2007}{\natexlab{a}}).

\bibitem[{\citenamefont{Oberbeck}(1879)}]{obe79}
\bibinfo{author}{\bibfnamefont{A.}~\bibnamefont{Oberbeck}},
  \bibinfo{journal}{Ann. Phys. Chem.} \textbf{\bibinfo{volume}{7}},
  \bibinfo{pages}{271} (\bibinfo{year}{1879}).

\bibitem[{\citenamefont{Boussinesq}(1903)}]{bou03}
\bibinfo{author}{\bibfnamefont{J.}~\bibnamefont{Boussinesq}},
  \emph{\bibinfo{title}{Theorie analytique de la chaleur, Vol. 2}}
  (\bibinfo{publisher}{Gauthier-Villars}, \bibinfo{address}{Paris},
  \bibinfo{year}{1903}).

\bibitem[{\citenamefont{Landau and Lifshitz}(1987)}]{ll87}
\bibinfo{author}{\bibfnamefont{L.~D.} \bibnamefont{Landau}} \bibnamefont{and}
  \bibinfo{author}{\bibfnamefont{E.~M.} \bibnamefont{Lifshitz}},
  \emph{\bibinfo{title}{Fluid Mechanics}} (\bibinfo{publisher}{Pergamon Press},
  \bibinfo{address}{Oxford}, \bibinfo{year}{1987}).

\bibitem[{\citenamefont{Chandrasekhar}(1981)}]{cha81}
\bibinfo{author}{\bibfnamefont{S.}~\bibnamefont{Chandrasekhar}},
  \emph{\bibinfo{title}{Hydrodynamic and Hydromagnetic Stability}}
  (\bibinfo{publisher}{Dover}, \bibinfo{address}{New York},
  \bibinfo{year}{1981}).

\bibitem[{\citenamefont{Ahlers et~al.}(2006{\natexlab{a}})\citenamefont{Ahlers,
  Brown, {{Fontenele Araujo}}, Funfschilling, Grossmann, and Lohse}}]{ahl06}
\bibinfo{author}{\bibfnamefont{G.}~\bibnamefont{Ahlers}},
  \bibinfo{author}{\bibfnamefont{E.}~\bibnamefont{Brown}},
  \bibinfo{author}{\bibfnamefont{F.}~\bibnamefont{{{Fontenele Araujo}}}},
  \bibinfo{author}{\bibfnamefont{D.}~\bibnamefont{Funfschilling}},
  \bibinfo{author}{\bibfnamefont{S.}~\bibnamefont{Grossmann}},
  \bibnamefont{and} \bibinfo{author}{\bibfnamefont{D.}~\bibnamefont{Lohse}},
  \bibinfo{journal}{J. Fluid Mech.} \textbf{\bibinfo{volume}{569}},
  \bibinfo{pages}{409} (\bibinfo{year}{2006}{\natexlab{a}}).

\bibitem[{\citenamefont{Zhang et~al.}(1997)\citenamefont{Zhang, Childress, and
  Libchaber}}]{zha97}
\bibinfo{author}{\bibfnamefont{J.}~\bibnamefont{Zhang}},
  \bibinfo{author}{\bibfnamefont{S.}~\bibnamefont{Childress}},
  \bibnamefont{and}
  \bibinfo{author}{\bibfnamefont{A.}~\bibnamefont{Libchaber}},
  \bibinfo{journal}{Phys. Fluids} \textbf{\bibinfo{volume}{9}},
  \bibinfo{pages}{1034} (\bibinfo{year}{1997}).
  
\bibitem[{\citenamefont{Sugiyama et~al.}(2007)\citenamefont{Sugiyama,
  Calzavarini, Grossmann, and Lohse}}]{sug07}
\bibinfo{author}{\bibfnamefont{K.}~\bibnamefont{Sugiyama}},
  \bibinfo{author}{\bibfnamefont{E.}~\bibnamefont{Calzavarini}},
  \bibinfo{author}{\bibfnamefont{S.}~\bibnamefont{Grossmann}},
  \bibnamefont{and} \bibinfo{author}{\bibfnamefont{D.}~\bibnamefont{Lohse}},
  \bibinfo{journal}{Europhys. Lett.} \textbf{\bibinfo{volume}{80}}, \bibinfo{pages}{34002}
  (\bibinfo{year}{2007}).  

\bibitem[{\citenamefont{Ahlers et~al.}(2007)\citenamefont{Ahlers, {{Fontenele
  Araujo}}, Funfschilling, Grossmann, and Lohse}}]{ahl07}
\bibinfo{author}{\bibfnamefont{G.}~\bibnamefont{Ahlers}},
  \bibinfo{author}{\bibfnamefont{F.}~\bibnamefont{{{Fontenele Araujo}}}},
  \bibinfo{author}{\bibfnamefont{D.}~\bibnamefont{Funfschilling}},
  \bibinfo{author}{\bibfnamefont{S.}~\bibnamefont{Grossmann}},
  \bibnamefont{and} \bibinfo{author}{\bibfnamefont{D.}~\bibnamefont{Lohse}},
  \bibinfo{journal}{Phys. Rev. Lett.} \textbf{\bibinfo{volume}{98}},
  \bibinfo{pages}{054501} (\bibinfo{year}{2007}).

\bibitem[{\citenamefont{Friend et~al.}(1991)\citenamefont{Friend, Ingham, and
  Ely}}]{fri91b}
\bibinfo{author}{\bibfnamefont{D.~G.} \bibnamefont{Friend}},
  \bibinfo{author}{\bibfnamefont{H.}~\bibnamefont{Ingham}}, \bibnamefont{and}
  \bibinfo{author}{\bibfnamefont{J.~F.} \bibnamefont{Ely}},
  \bibinfo{journal}{J. Phys. Chem. Ref. Data} \textbf{\bibinfo{volume}{20}},
  \bibinfo{pages}{275} (\bibinfo{year}{1991}).

\bibitem[{\citenamefont{Ciliberto et~al.}(1996)\citenamefont{Ciliberto, Cioni,
  and Laroche}}]{cil96}
\bibinfo{author}{\bibfnamefont{S.}~\bibnamefont{Ciliberto}},
  \bibinfo{author}{\bibfnamefont{S.}~\bibnamefont{Cioni}}, \bibnamefont{and}
  \bibinfo{author}{\bibfnamefont{C.}~\bibnamefont{Laroche}},
  \bibinfo{journal}{Phys. Rev. E} \textbf{\bibinfo{volume}{54}},
  \bibinfo{pages}{R5901} (\bibinfo{year}{1996}).

\bibitem[{\citenamefont{Ahlers et~al.}(2006{\natexlab{b}})\citenamefont{Ahlers,
  Brown, and Nikolaenko}}]{ahl06b}
\bibinfo{author}{\bibfnamefont{G.}~\bibnamefont{Ahlers}},
  \bibinfo{author}{\bibfnamefont{E.}~\bibnamefont{Brown}}, \bibnamefont{and}
  \bibinfo{author}{\bibfnamefont{A.}~\bibnamefont{Nikolaenko}},
  \bibinfo{journal}{J. Fluid Mech.} \textbf{\bibinfo{volume}{557}},
  \bibinfo{pages}{347} (\bibinfo{year}{2006}{\natexlab{b}}).

\bibitem[{\citenamefont{Schlichting and Gersten}(2000)}]{sch00}
\bibinfo{author}{\bibfnamefont{H.}~\bibnamefont{Schlichting}} \bibnamefont{and}
  \bibinfo{author}{\bibfnamefont{K.}~\bibnamefont{Gersten}},
  \emph{\bibinfo{title}{Boundary layer theory}} (\bibinfo{publisher}{Springer
  Verlag}, \bibinfo{address}{Berlin}, \bibinfo{year}{2000}),
  \bibinfo{edition}{8th} ed.

\bibitem[{\citenamefont{Stewartson}(1964)}]{ste64}
\bibinfo{author}{\bibfnamefont{K.}~\bibnamefont{Stewartson}},
  \emph{\bibinfo{title}{The theory of laminar boundary layers in compressible
  fluids}} (\bibinfo{publisher}{Oxford University Press},
  \bibinfo{year}{1964}).

\bibitem[{\citenamefont{Krishnamurti and Howard}(1981)}]{kri81}
\bibinfo{author}{\bibfnamefont{R.}~\bibnamefont{Krishnamurti}}
  \bibnamefont{and} \bibinfo{author}{\bibfnamefont{L.~N.}
  \bibnamefont{Howard}}, \bibinfo{journal}{Proc. Natl. Acad. Sci.}
  \textbf{\bibinfo{volume}{78}}, \bibinfo{pages}{1981} (\bibinfo{year}{1981}).

\bibitem[{\citenamefont{Press et~al.}(1986)\citenamefont{Press, Teukolsky,
  Vetterling, and Flannery}}]{pre86}
\bibinfo{author}{\bibfnamefont{W.}~\bibnamefont{Press}},
  \bibinfo{author}{\bibfnamefont{S.}~\bibnamefont{Teukolsky}},
  \bibinfo{author}{\bibfnamefont{W.}~\bibnamefont{Vetterling}},
  \bibnamefont{and} \bibinfo{author}{\bibfnamefont{B.}~\bibnamefont{Flannery}},
  \emph{\bibinfo{title}{Numerical Recipes}} (\bibinfo{publisher}{Cambridge
  University Press}, \bibinfo{address}{Cambridge}, \bibinfo{year}{1986}).


\bibitem[Ahlers {\em et~al.\/}(1994)]{ACBS94}\bibnamefont{G. Ahlers, D. S.
 Cannell,  L. I. Berge, and  S. Sakurai},
{\em Phys. Rev. E\/} {\bf 49}, 545-553 (1994).

\bibitem[{\citenamefont{M\"uller et~al.}(1976)\citenamefont{M\"uller, Ahlers,
  and Pobell}}]{mue76}
\bibinfo{author}{\bibfnamefont{K.}~\bibnamefont{M\"uller}},
  \bibinfo{author}{\bibfnamefont{G.}~\bibnamefont{Ahlers}}, \bibnamefont{and}
  \bibinfo{author}{\bibfnamefont{F.}~\bibnamefont{Pobell}},
  \bibinfo{journal}{Phys. Rev. B} \textbf{\bibinfo{volume}{14}},
  \bibinfo{pages}{2096} (\bibinfo{year}{1976}).

\bibitem[{\citenamefont{Ahlers}(2000)}]{ahl00}
\bibinfo{author}{\bibfnamefont{G.}~\bibnamefont{Ahlers}},
  \bibinfo{journal}{Phys. Rev. E} \textbf{\bibinfo{volume}{63}},
  \bibinfo{pages}{015303} (\bibinfo{year}{2000}).

\bibitem[{\citenamefont{Roche et~al.}(2001)\citenamefont{Roche, Castaing,
  Chabaud, Hebral, and Sommeria}}]{RCCHS01}
\bibinfo{author}{\bibfnamefont{P.}~\bibnamefont{Roche}},
  \bibinfo{author}{\bibfnamefont{B.}~\bibnamefont{Castaing}},
  \bibinfo{author}{\bibfnamefont{B.}~\bibnamefont{Chabaud}},
  \bibinfo{author}{\bibfnamefont{B.}~\bibnamefont{Hebral}}, \bibnamefont{and}
  \bibinfo{author}{\bibfnamefont{J.}~\bibnamefont{Sommeria}},
  \bibinfo{journal}{Eur. Phys. J. B} \textbf{\bibinfo{volume}{24}},
  \bibinfo{pages}{405} (\bibinfo{year}{2001}).

\bibitem[{\citenamefont{Brown et~al.}(2005{\natexlab{b}})\citenamefont{Brown,
  Funfschilling, Nikolaenko, and Ahlers}}]{bro05}
\bibinfo{author}{\bibfnamefont{E.}~\bibnamefont{Brown}},
  \bibinfo{author}{\bibfnamefont{D.}~\bibnamefont{Funfschilling}},
  \bibinfo{author}{\bibfnamefont{A.}~\bibnamefont{Nikolaenko}},
  \bibnamefont{and} \bibinfo{author}{\bibfnamefont{G.}~\bibnamefont{Ahlers}},
  \bibinfo{journal}{Phys. Fluids} \textbf{\bibinfo{volume}{17}},
  \bibinfo{pages}{075108} (\bibinfo{year}{2005}{\natexlab{b}}).

\bibitem[{\citenamefont{Brown and Ahlers}(2007{\natexlab{b}})}]{bro07b}
\bibinfo{author}{\bibfnamefont{E.}~\bibnamefont{Brown}} \bibnamefont{and}
  \bibinfo{author}{\bibfnamefont{G.}~\bibnamefont{Ahlers}},
  \bibinfo{journal}{EPL} \textbf{\bibinfo{volume}{80}}, \bibinfo{pages}{14001}
  (\bibinfo{year}{2007}{\natexlab{b}}).

\bibitem[{\citenamefont{Sugiyama et~al.}(2008)\citenamefont{Sugiyama,
  Calzavarini, Grossmann, and Lohse}}]{sug08}
\bibinfo{author}{\bibfnamefont{K.}~\bibnamefont{Sugiyama}},
  \bibinfo{author}{\bibfnamefont{E.}~\bibnamefont{Calzavarini}},
  \bibinfo{author}{\bibfnamefont{S.}~\bibnamefont{Grossmann}},
  \bibnamefont{and} \bibinfo{author}{\bibfnamefont{D.}~\bibnamefont{Lohse}},
  \bibinfo{journal}{J. Fluid Mech., to be 
submitted} \textbf{\bibinfo{volume}{}}
  \bibinfo{pages}{} (\bibinfo{year}{2008}).

\end{thebibliography}

\end{document}